# Spatial mapping and analysis of graphene nanomechanical resonator networks

*Brittany Carter[1,2,3], Viva R. Horowitz[4], Uriel Hernandez[1,2,3], David J. Miller[1,2,3], Andrew Blaikie[1,2,3], and Benjamín J. Alemán[1,2,3,5,]\**

[1]Department of Physics, University of Oregon, Eugene, Oregon, 97403, United States
[2]Materials Science Institute, University of Oregon, Eugene, Oregon, 97403, United States
[3]Center for Optical, Molecular, and Quantum Science, University of Oregon, Eugene, Oregon, 97403, United States
[4]Physics Department, Hamilton College, Clinton, New York, 13323, United States
[5]Phil and Penny Knight Campus for Accelerating Scientific Impact, University of Oregon, Eugene, Oregon, 97403, United States

*Corresponding Author: Benjamín J. Alemán, baleman@uoregon.edu

**Nanoelectromechanical (NEMS) resonator networks have drawn increasing interest due to their potential applications in emergent behavior, sensing, phononics, and mechanical information processing. A challenge toward realizing these large-scale networks is the ability to controllably tune and reconfigure the collective, macroscopic properties of the network, which relies directly on the development of methods to characterize the constituent NEMS resonator building blocks and their coupling. In this work, we demonstrate a scalable optical technique to spatially map graphene NEMS networks and read out the fixed-frequency collective response as a single vector. Using the response vectors, we introduce an efficient algebraic approach to quantify the site-specific elasticity, mass, damping, and coupling parameters of network clusters. We apply this technique to accurately characterize single uncoupled resonators and coupled resonator pairs by sampling them at just two frequencies, and without the use of curve fitting or the associated *a priori* parameter estimates. Our technique may be applied to a range of classical and quantum resonator systems and fills in a vital gap for programming NEMS networks.**

Assemblies of interacting resonators are abundant in nature and are found in a growing number of synthetic many-body systems including trapped ion chains[1], superconducting qubit arrays[2], and optical lattices[3]. A compelling testbed for these assemblies is the programmable network of nanoelectromechanical systems (NEMS)[4,5]—wherein the mechanical properties (*e.g.* resonance frequency) and coupling of constituent NEMS resonators are finely tuned in order to tailor the collective properties of the network. Once realized, these on-chip programmable systems will unlock powerful applications such as reconfigurable phononic crystals and waveguides[6,7] tunable thermal transport[5,8], and mechanics-based circuits[9], computing[4,10] and simulation[9].

Recent advancements towards realizing NEMS networks include demonstrations of collective phenomena in small-scale modular assemblies[9] and lattices[4,5] and the development of basic programming tools to tune the resonance frequency of individual resonators[11] and the coupling between single pairs[12,13,14]. Despite these successes, there persists a fundamental need for scalable, spatially resolved characterization methods to read out the mechanical properties of the network resonator building blocks and the linkages between them, methods needed to ascertain the spatial configuration of the resonators and to initialize and correct the network to a desired state.

NEMS characterization is often performed by analyzing resonance peaks in displacement spectra. For example, amplitude spectra near resonance are used to quantify the dissipation and eigenfrequency of single resonators[11] and to ascertain coupling strengths between resonators[13,14,15]. However, as the size of the network is expanded—even modestly—it becomes increasingly challenging to obtain site-specific mechanical information from spectra alone. Notably, spectral peaks corresponding to a hybridized mode[16] cannot be used to distinguish which resonators participate in the mode, thereby frustrating efforts to characterize and tune specific resonators. Moreover, weak coupling—an essential feature of many collective phenomena[9,17,18]—can be impossible to detect spectroscopically because of its small signal and indistinguishable mode splitting[19], while spurious non-mechanical spectral features often confound characterization attempts[9,20,21]. In addition, network parameters inferred from spectral data using traditional analysis approaches (*e.g.* non-linear least square) require *a priori* knowledge of the network parameters as input guesses to achieve reasonable model predictions. Even so, correlations and metastable least-squares minima often lead to large uncertainty in parameter estimates[22].

Here we demonstrate a site-specific method to quantify the elasticity, mass, damping, and coupling of individual resonators in a nanomechanical network that overcomes the limitations of current approaches. In our method, which we call NetMAP (Network Mapping and Analysis of Parameters), we employ optical scanning interference microscopy[5,23,24,25] (SIM) to spatially image the amplitude and phase response of hybridized vibrational modes in a network of graphene membrane nanomechanical resonators. The resulting SIM images provide the spatial address for each resonator, which we use to measure fixed-frequency response vectors. Using just two vectors, we solve the reciprocal-space algebraic form of the network's equation of motion to determine all the mechanical parameters of the network, even in the case of weak coupling and without any *a priori* knowledge (*e.g.* parameter estimates). The combination of our graphene resonator arrays, NetMAP, and existing methods to tune resonators[11] establishes a viable platform for programmable nanomechanical networks.

In our approach, we model the resonator network as a linear chain of masses and springs, depicted in **Figure 1a.** The system of equations describing this model can be organized into matrix form

$$\boldsymbol{M}\ddot{\vec{x}} + \mathbf{B}\dot{\vec{x}} + \boldsymbol{K}\vec{x} = \vec{F}. \tag{1}$$

where $\boldsymbol{M}$, $\boldsymbol{B}$, and $\boldsymbol{K}$, are the mass, damping, and elasticity matrices, respectively. Using $x_n(t) = |Z_n(\omega)|e^{i(\omega t - \phi_n(\omega))}$ as the response of the $n^{th}$ resonator, we can write Eq. (1) in steady-state form:

$$\boldsymbol{\mathcal{M}}(\omega)\vec{Z}(\omega) = \vec{F} \tag{2}$$

Here $\omega$ is the drive frequency of $\vec{F}$, $\boldsymbol{\mathcal{M}}(\omega) \equiv -\omega^2 \boldsymbol{M} + i\omega \boldsymbol{B} + \boldsymbol{K}$, and $\vec{Z}(\omega)$ is a vector of the complex responses of each resonator in the network, $Z_n(\omega)$. Our approach determines $\boldsymbol{M}$, $\boldsymbol{B}$, and $\boldsymbol{K}$ by measuring the response vector $\vec{Z}(\omega)$. For finite clusters of size $N$, $\boldsymbol{\mathcal{M}}(\omega)$ for a given $\omega$ has $4N$ unknown parameters, and for each measurement of $\vec{Z}(\omega)$, Eq. (2) provides $2N$ equations. Thus, to determine the unknown parameters of $\boldsymbol{M}$, $\boldsymbol{B}$, and $\boldsymbol{K}$, $\vec{Z}(\omega)$ must be measured at a minimum of two drive frequencies, $\omega_a$ and $\omega_b$ (See SI Discussion). We combine and reorganize the equations corresponding to $\vec{Z}(\omega_a)$ and $\vec{Z}(\omega_b)$ in Eq. (2) to obtain

$$\boldsymbol{\mathcal{Z}}\vec{p} = \vec{0} \tag{3}$$

where $\mathcal{Z}$ is a $4N \times 4N$ real-valued matrix of known coefficients determined by $\omega_a$, $\omega_b$, $\vec{Z}(\omega_a)$, and $\vec{Z}(\omega_b)$. The *parameters vector*—$\vec{p}$—is an $4N$-dimensional vector comprised of all the unknown elements of $\boldsymbol{M}$, $\boldsymbol{B}$, $\boldsymbol{K}$, and $\vec{F}$.

To solve Eq. (*3*) for $\vec{p}$, we determine the null space of $\mathcal{Z}$ via singular value decomposition (SVD). We use the `NumPy` SVD package in Python, which outputs normalized values of $\vec{p}$ that can be used to calculate any ratio of interest (*e.g.* the quality factor, coupling strength, etc.) and to plot the predicted $\vec{Z}(\omega)$ to validate against measurement. The normalized values in $\vec{p}$ can be un-normalized by separately determining one parameter.

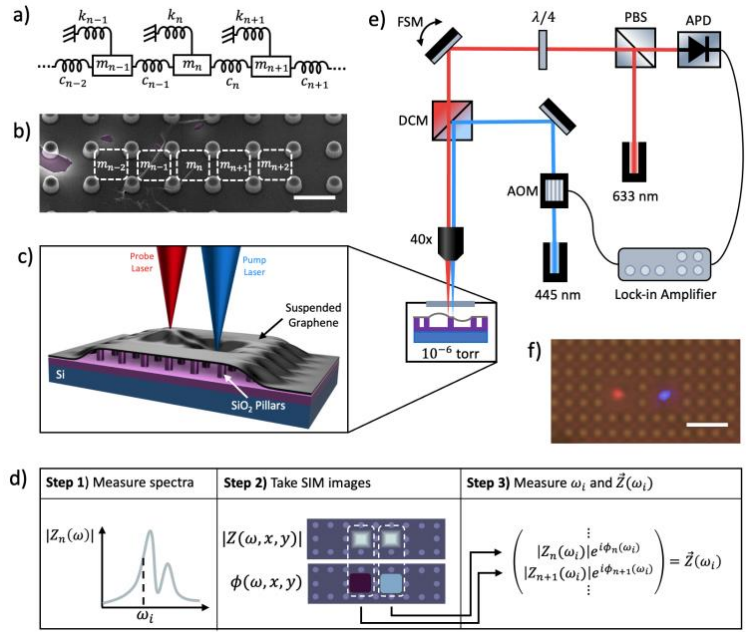

**Figure 1 a)** Linear mass and spring model showing intrinsic springs ($k_n$), coupling springs ($c_n$), and masses ($m_n$). **b)** SEM image of graphene suspended over pillars. Suspended regions between pillars are depicted as a linear chain of masses. Scale bar is $3\ \mu m$. **c)** Cross section view of suspended graphene device showing Si base with $SiO_2$ pillars and suspended graphene. The depicted pump and probe lasers are aligned to drive the right-side resonator and measure the motion of the left-side resonator. **d)** Steps for measuring $\vec{Z}(\omega_i)$ showing (1) an amplitude spectrum of the $n^{th}$ resonator, (2) spatial images of a cluster of two resonators, and (3) the complex response vector of the cluster when driven at $\omega_i$. **e)** Schematic of optical set up showing 445 nm pump laser modulated with an AOM and coupled into the optical path with a dichroic mirror (DCM). The 633 nm probe laser is deflected by a polarizing beam splitter (PBS) and polarized by a quarter waveplate ($\lambda/4$). The probe position is controlled with a fast-scanning mirror (FSM). The reflected probe passes back through the PBS, and the interference signal is detected by an avalanche photodiode (APD) and resolved by a lock-in amplifier. Both the pump and probe lasers are focused onto the sample through a 40x objective lens and the sample is under vacuum at $10^{-6}$ torr. **f)** Top-down optical image of the sample under vacuum, with pillars seen as small orange dots. Here, the pump (blue) and probe (red) beams are focused onto neighboring resonator regions of suspended graphene, with a scalebar of 6 μm.

We demonstrate NetMAP on network clusters of graphene resonators suspended over pillar arrays. We created the arrays by patterning SiO$_2$/Si substrates using e-beam lithography followed by a dry reactive ion etch (CHF$_4$), resulting in SiO$_2$ pillars $\sim 600$ nm in height (see Si for details). We then suspended commercially grown CVD graphene over the pillar arrays using a wet transfer method[26] (see Si for details), resulting in an array of interconnected, suspended graphene resonators, shown in one configuration in **Figure 1b**. Additionally, we omit pillars throughout the array to create larger-size resonators and resonator pairs (**Figure 1c,e**). The lateral size of the membrane resonators varies from $3 - 6$ µm. The resonators are directly connected by suspended graphene, which provides a mechanism for elastic coupling represented by the coupling spring constants, $c_n$, in **Figure 1a**.

With the goal of determining $\vec{p}$ for a local cluster of suspended graphene resonators, we must first construct the matrix $\mathcal{Z}$, which we achieve in three steps (**Figure 1d**): (1) Determine the drive frequency $\omega$ that provides large signal-to-noise ratio (SNR) for $|Z_n(\omega)|$, (2) quantify the size and spatial configuration of the cluster, and (3) measure $\vec{Z}(\omega_a)$, $\vec{Z}(\omega_b)$ and use $\omega_a, \omega_b, \vec{Z}(\omega_a), \vec{Z}(\omega_b)$ to construct $\mathcal{Z}$.

The first step is to determine which drive frequencies $\omega$ provide a large response signal $|Z_n(\omega)|$, which we achieve by measuring amplitude and phase spectra of the $n^{th}$ resonator and searching for resonance peaks (**Figure 1d, Step 1**). To measure the spectra, we position a focused optical pump (445 nm) and probe (633 nm) on the $n^{th}$ resonator. While modulating the pump with an AOM, we measure the corresponding amplitude and phase via interferometry and lock-in amplification (see **Figure 1c,e,f**). All measurements are taken in vacuum ($\sim 10^{-7}\ torr$) at room temperature. From the resulting spectra, we locate resonance peaks to determine a range of frequencies that provide a large $|Z_n(\omega)|$. To further increase the signal $|Z_n(\omega)|$, we increase the amplitude of the pump laser until $|Z_n(\omega)|$ is just below the limit of the linear regime. For all further measurements and steps, we fix the coordinates of the pump at this initial position.

Second, we quantify the size $N$ and spatial configuration of the cluster by obtaining images of the mechanical motion of the area surrounding the driven resonator via scanning interferometric microscopy (SIM)[23]. For SIM, we modulate the fixed pump at a value of $\omega$, chosen to correspond to a large signal in the measured amplitude spectrum. We then raster the probe across the sample while recording both amplitude and phase, resulting in two-dimensional spatial images $|Z(\omega, x, y)|$ and $\phi(\omega, x, y)$ (**Figure 1d, Step 2**). SIM images are taken over a large enough area to characterize the local vicinity of the driven resonator, typically about 20 µm $\times$ 20 µm in size. To quantify $N$, we use $|Z(\omega, x, y)|$ and $\phi(\omega, x, y)$ to count the number of resonators in the cluster, corresponding to the number of regions that have local amplitude maxima and constant phase. To confirm $N$ and map the spatial configuration of the cluster, we cross-correlate $|Z(\omega, x, y)|$ with an optical brightfield (**Figure 1f**) and SEM (**Figure 1b**) image to match peak amplitudes with specific areas of the suspended membrane. Our home-built SIM software allows for easy point-and-click positioning of the probe, which we use to collect spectra (step one) for each resonator in the cluster. With a full set of spectra, we determine which frequencies $\omega$ result in a largest overall response signal. Moreover, we use the point-and-click positioning feature to finely tune the $x$-$y$ probe coordinates over each resonator to further maximize the signal.

The third and final step is to measure the components of $\mathcal{Z}$ — constructed with $\omega_a, \omega_b, \vec{Z}(\omega_a), \vec{Z}(\omega_b)$. To obtain the first set $\left(\omega_a, \vec{Z}(\omega_a)\right)$, we position the probe on the $n^{th}$ resonator, but rather than fixing $\omega_a$ and measuring $Z_n(\omega_a)$, we use a phase-locked loop (PLL) to monitor the time series of frequency and amplitude, $\Omega_n(\omega_a)$ and $A_n(\omega_a)$, for a fixed phase input of $\phi_n(\omega_a)$. The phase input is determined by the value of the phase spectrum at the target $\omega_a$. We record each time series until the amplitude SNR reaches a predetermined value of $\sim 10^3$, typically resulting in $\sim 10^4$ discrete measurements. We define the SNR as the ratio of the mean of the series to the standard error ($|\bar{A}|/\sigma_{\bar{A}}$). Although the best choice of a target $\omega_a$ for high SNR would be at the peak resonance, we chose frequencies slightly off resonance to reduce error from frequency-dependent linear phase lags intrinsic to the optical set up (See SI for details). Additionally, the phase must be corrected for this lag prior to calculating $Z_n(\omega_a)$. We correct the phase by linearly fitting a region of the phase spectrum far off resonance to determine an intercept, $\phi_0$, and the time delay slope, $\tau$. We then subtract the quantity $\phi_0 - \omega_i \tau$ from the PLL input phase (See SI for details). After repeating this PLL measurement for all $N$ resonators in the cluster, we calculate $(\omega_a, \vec{Z}(\omega_a))$ as $\omega_a = \sum_{n=1}^{N} \frac{\bar{\Omega}_n(\omega_a)}{N}$ and $\vec{Z}(\omega_a) = \{\overline{A_1}e^{i\phi_1(\omega_a)}, \dots, \overline{A_N}e^{i\phi_N(\omega_a)}\}$, in which we use the corrected phase values, $\phi_n(\omega_a)$, and the time series averages, $\bar{\Omega}_n(\omega_a)$ and $\overline{A_n}(\omega_a)$. We repeat this procedure to calculate $(\omega_b, \vec{Z}(\omega_b))$, completing all the measurements needed to calculate the components of $\mathcal{Z}$.

As a first application of NetMAP, we determined the mechanical parameters of the simplest network cluster of size $N = 1$, corresponding to a single, uncoupled resonator. We first position the pump and probe over the resonator region highlighted in **Figure 2a,** and acquire amplitude and phase spectra. The resulting spectra revealed a single peak in the amplitude, shown as grey data points in **Figure 2f (upper)**, that corresponded to a corrected phase of $\pi/2$ (**Figure 2f (lower)**, grey), consistent with a single, uncoupled resonator. To confirm the size of this cluster was $N = 1$, we took a SIM scan at 18.81 $MHz$, resulting in the amplitude and phase spatial maps shown in **Figure 2b.** In the amplitude map, we observed an amplitude maximum ($\sim 10^{-4}$ mV) within one localized $\sim 6 \times 6 \; \mu m^2$ region of the suspended graphene, which matched the size and location of the region highlighted in **Figure 2a**. The spatial undulations in the amplitude and phase near the edge of the resonator region are likely due to interactions with the pillars. Outside of the resonator region, the amplitude decreases by more than two orders of magnitude ($\sim 10^{-6}$ mV). Moreover, the resonator region has nearly constant phase ($STD = 0.07$ rad), implying it moves in unison, as expected for the fundamental mode of a single resonator. Away from the resonator, the phase is noisier ($F_0 \sim 74$, $p \sim 10^{-11}$), with an increase in the standard deviation by over an order of magnitude ($STD = 0.6$ rad). Lastly, as seen in the line scans (**Figure 2b,c**), the amplitude has one solitary lobe with constant phase. Altogether, we conclude that this local cluster consists of a single, uncoupled resonator.

Given our single-resonator cluster, we next obtained $\omega_a, \omega_b, \vec{Z}(\omega_a)$ and $\vec{Z}(\omega_b)$, needed to compute $\mathcal{Z}$. We chose $\omega_a$ and $\omega_b$ to be on each side of the resonance peak, corresponding to corrected phase values of $\phi_1(\omega_a) = -0.33 \pm 0.12$ rad and $\phi_1(\omega_b) = -2.78 \pm 0.12$ rad. We completed the PLL measurement when the amplitude SNR reached $> 5000$. The PLL time series of amplitude and frequency are shown as 2D boxplots for a phase lock of $\phi_1(\omega_a)$ in Figure 2d and for $\phi_1(\omega_b)$ in **Figure 2e**. The mean values of amplitude, $\overline{A_1}(\omega_a)$ and $\overline{A_1}(\omega_b)$, and frequency, $\overline{\Omega_1}(\omega_a)/2\pi$ and $\overline{\Omega_1}(\omega_b)/2\pi$, are plotted as diamond points in **Figure 2f (upper)**, with the color of each point corresponding to the box-and-whisker plots (**Figure 2d,e**). Errors are calculated standard error. The corrected phase values, $\phi_1(\omega_a)$ and $\phi_1(\omega_b)$, and mean values of frequency, $\overline{\Omega_1}(\omega_a)/2\pi$ and $\overline{\Omega_1}(\omega_b)/2\pi$, are plotted as diamond points in **Figure 2f (lower).** Using these measurements, we obtain $\omega_a = \overline{\Omega_1}(\omega_a)$, $\omega_b = \overline{\Omega_1}(\omega_b)$, $\vec{Z}(\omega_a) = \{\overline{A_1}(\omega_a)e^{i\phi_1(\omega_a)}\}$, and $\vec{Z}(\omega_b) = \{\overline{A_1}(\omega_b)e^{i\phi_1(\omega_b)}\}$.

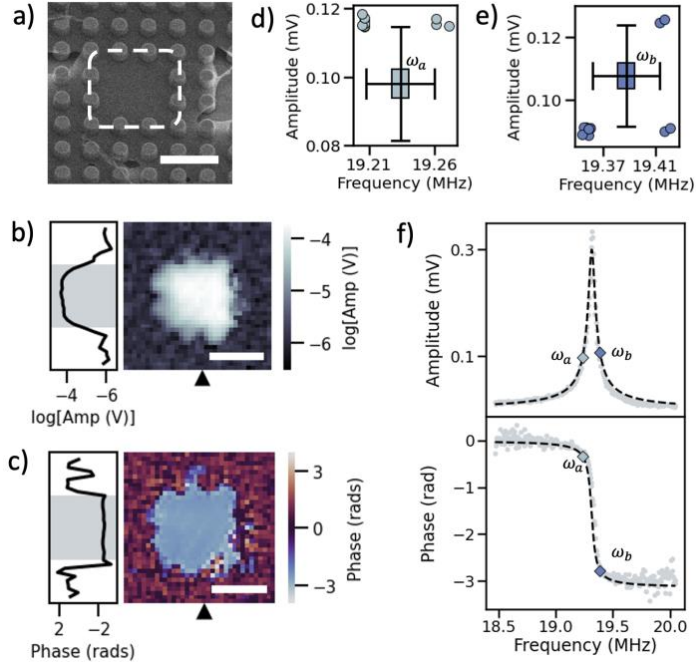

**Figure 2 a)** SEM of driven uncoupled resonator, with pillar radii of 0.5 μm and pillar pitch 2 μm. Scale bar is 4 μm. **b)** Amplitude and **c)** phase spatial maps at a drive frequency of 18.81 MHz, scale bars are 5 μm. Phase spatial map shows uncorrected wrapped phase values. Black triangles on the bottom axis indicate the location of the vertical line scan on the left-side axis. 2D boxplot of PLL measurement distribution of frequency and amplitude for **d)** phase lock of $\phi_1(\omega_a) = 1.84$ rad and **e)** phase lock of $\phi_2(\omega_b) = -1.28$ rad. 2D boxplots shows median, upper, and lower quartile ranges with whiskers that extend to include 1.5 IQR. Plotted circles represent datapoints that were outliers in both frequency and amplitude. **f)** Amplitude and corrected phase spectra of driven resonator. The diamond points in the amplitude spectrum (upper) correspond to the mean PLL measurements of amplitude ($\overline{A_1}(\omega_a) = 0.09827 \pm 0.00002$ mV and $\overline{A_1}(\omega_b) = 0.10758 \pm 0.00002$ mV) and frequency ($\overline{\Omega_1}(\omega_a)/2\pi = 19.2337 \pm 0.00003$ MHz and $\overline{\Omega_1}(\omega_b)/2\pi = 19.3876 \pm 0.00004$ MHz), where the uncertainties are standard error. The diamond points in the phase spectrum (lower) correspond to the locked phase values, $\phi_1(\omega_a)$ and $\phi_1(\omega_b)$, and the mean frequency values, $\overline{\Omega_1}(\omega_a)/2\pi$ and $\overline{\Omega_1}(\omega_b)/2\pi$. The black dotted line represents $|Z_1(\omega)|$ and $\phi_1(\omega)$ generated from the normalized $\vec{p}$.

To solve for the parameters vector, $\vec{p}$, we populate matrix $\mathcal{Z}$ with coefficients of the experimentally measured $\omega_a, \omega_b, \vec{Z}(\omega_a)$ and $\vec{Z}(\omega_b)$. For a single-resonator cluster, the reduced system of equations, $\mathcal{Z}\vec{p} = \vec{0}$, is

$$\begin{pmatrix} -\omega_a^2 Re[Z_1(\omega_a)] & -\omega_a Im[Z_1(\omega_a)] & Re[Z_1(\omega_a)] & -1 \\ -\omega_a^2 Im[Z_1(\omega_a)] & \omega_a Re[Z_1(\omega_a)] & Im[Z_1(\omega_a)] & 0 \\ -\omega_b^2 Re[Z_1(\omega_b)] & -\omega_b Im[Z_1(\omega_b)] & Re[Z_1(\omega_b)] & -1 \\ -\omega_b^2 Im[Z_1(\omega_b)] & \omega_b Re[Z_1(\omega_b)] & Im[Z_1(\omega_b)] & 0 \end{pmatrix} \begin{pmatrix} m \\ b \\ k \\ f \end{pmatrix} = \begin{pmatrix} 0 \\ 0 \\ 0 \\ 0 \end{pmatrix} \quad (4)$$

Given $N = 1$, $\vec{p}$ has four unknowns: $m, b, k$, and $f$. We applied SVD to solve Eq (4) for $\vec{p}$, with the normalized values of $\vec{p}$ and associated errors listed in **Table 1**. We propagated the errors in $\vec{p}$ by sampling from Gaussian mean distributions of the two phases, mean drive frequencies, and mean amplitudes to generate hundreds of variations of $\mathcal{Z}$. We then used SVD to solve for $\vec{p}$ corresponding to each $\mathcal{Z}$, resulting in distributions of each output parameter (see SI for details).

To assess how accurately the resulting $\vec{p}$ characterized the single resonator, we compared the normalized values to expected values. The spring constant of suspended multi-layer graphene[27] is $\sim 1-5$ N/m, so normalizing $\vec{p}$ by $k_1 = 1$ N/m provides order-of-magnitude estimates of all other parameters. We can estimate the mass of the resonator by using the area density of pristine graphene ($\rho = 0.75$ mg/m$^2$) and an approximated area based on the suspended region highlighted in **Figure 2a**, which gives a mass of $2.7 \times 10^{-17}$ kg. While this mass estimate is lower than the value in **Table 1**, graphene contamination can increase the mass[28,29] by $\sim 10 \times$, putting the predicted $m$ within the expected range. In addition, we estimated the damping of the resonator by fitting the amplitude spectra to find the full width half max ($FWHM$) and the center frequency ($\omega_0$) (see SI Discussion 4). Using the fitted values, along with an estimation of $k = 1$ N/m, we estimate the damping to be $b = 1.92 \times 10^{-11}$ kg/s, which agrees with the value in **Table 1**.

To evaluate the predictive power of the NetMAP results, we used $\vec{p}$ to calculate $|Z_1(\omega)|$ and $\phi_1(\omega)$ and compared the results to the experimental spectra. The $\vec{p}$-calculated $|Z_1(\omega)|$ and $\phi_1(\omega)$ are shown as black dashed lines overlayed on the spectra (**Figure 2f)** and the resulting $R^2$ values are listed in Table 1. We see that, despite building the analytical spectra from just two data points, the prediction agrees well with the data, accounting for a minimum of 97% of variation. Therefore, the normalized $\vec{p}$ can predict the cluster response exceptionally well across the tested frequency range.

As an additional validation, we compared the NetMAP $\vec{p}$ to that obtained from non-linear least-squares fitting with order of magnitude initial guesses (Unity LS), which is perhaps the most common means of characterizing resonant systems. The fit parameter estimates obtained using `Lmfit` in Python are listed in **Table 1**. To compare the predicted $\vec{p}$ from each method, we employ a two-tailed t-test with the Unity LS parameters as reference. We find the damping $b$ ($p = 0.87$) agrees between both approaches but the mass $m$ ($p = 0.004$) and force $F$ ($p = 0.03$) do not agree. However, the mass from Unity LS still falls within the expected range and the force only differs by 7%. We also find that the $\vec{p}$ from Unity LS had similar predictive power when compared with the experimental spectra (see **Table 1**). Thus, we conclude that both methods are comparable for characterizing this system of a single resonator cluster. Furthermore, we used NetMAP to characterize an additional $N = 1$ cluster and attained similar results (see SI).

| Mechanical Parameter | NetMAP | Unity LS |
|---|---|---|
| $k$ [N/m] | 1 | 1 |
| $m$ [$10^{-17}$ kg] | $6.790 \pm 0.002$ | 6.785 |
| $b$ [$10^{-11}$ kg/s] | $2.388 \pm 0.643$ | 2.283 |
| $F$ [$10^{-7}$ au] | $8.713 \pm 0.286$ | 8.107 |
| Amplitude $R^2$ | 0.98 | 0.97 |
| Phase $R^2$ | 0.97 | 0.99 |

**Table 1**: Normalized values of $\vec{p}$ from NetMAP and Unity LS for N=1 cluster

We next test NetMAP on a system with more degrees of freedom, on a cluster size of $N = 2$. To begin, we position the pump and probe over the resonator region highlighted as R1 in **Figure 3a** and acquire amplitude and phase spectra. The resulting amplitude spectrum (**Figure 3f (upper), grey data points)** revealed two closely spaced peaks, each corresponding to a corrected phase of $\sim \pi/2$ (**Figure 3f**

**(lower)**, grey), indicative of the two hybridized modes of a pair of coupled resonators[19]. To test if the spectral features correspond to a resonator pair, we took a SIM scan across the region at a frequency below the first peak ($f_1 = 21.51\ MHz$). The resulting spatial maps (**Figure 3b,c**) show two distinct high-intensity amplitude regions with nearly constant phase, as highlighted in the line profiles (**Figure 3f**). The first high-amplitude region was centered in a $6 \times 6\ \mu m^2$ region, which matched the size and location of the driven resonator R1 shown in **Figure 3a**. The second region matched the size and location of a $3 \times 3\ \mu m^2$ membrane highlighted as R2 in **Figure 3a**. The mean phases of R1 and R2 differed by 0.12 rad, or ~6.9°, ($p$~0.001), indicating they move in near unison, in accord with expectations for the symmetric mode of a coupled pair[23]. We search for the asymmetric mode with an additional SIM scan at a frequency above the second spectral peak ($f_2 = 22.55\ MHz$). In the resulting amplitude map and cross-section (**Figure 3d**), we see two regions with high amplitude situated at the same locations as R1 and R2 in **Figure 3a,b**. The phase map and profile (**Figure 3e**) reveal the phase in each region is relatively uniform, but the regions differ from each other by ~$\pi$ rad, as expected for the asymmetric mode. From the spectra and SIM data, we conclude that the R1 and R2 resonators form a coupled pair.

To measure $\omega_a, \omega_b, \vec{Z}(\omega_a)$ and $\vec{Z}(\omega_b)$ for the coupled pair, we first aligned the probe over R2 and took spectra (**Figure 3i**). Using the R1 and R2 spectra, we chose a target value of $\omega_a$ below the symmetric mode peak and $\omega_b$ above the antisymmetric mode peak. We then positioned the probe over R1 and acquired two PLL time-series measurements at corrected phase values of $\phi_1(\omega_a)$ and $\phi_1(\omega_b)$. We completed each PLL measurement when the amplitude SNR reached ~3000. The R1 PLL data are shown as 2D boxplots for phase lock values of $\phi_1(\omega_a)$ in **Figure 3f** and $\phi_1(\omega_b)$ in **Figure 3g**. The mean values of amplitude, ($\overline{A_1}(\omega_a)$ and $\overline{A_1}(\omega_b)$), and frequency ($\overline{\Omega_1}(\omega_a)/2\pi$ and $\overline{\Omega_1}(\omega_b)/2\pi$), are plotted as diamond points in **Figure 3j (upper)**, with the color of each point corresponding to the 2D boxplots (**Figure 3f,g**). The corrected phase values, $\phi_1(\omega_a)$ and $\phi_1(\omega_b)$, and mean values of frequency, $\overline{\Omega_1}(\omega_a)/2\pi$ and $\overline{\Omega_1}(\omega_b)/2\pi$, are plotted as diamond points in **Figure 3j (lower).** We repeated PLL measurements for the second resonator by placing the probe over R2, the pump fixed over R1, with corrected phase values of $\phi_2(\omega_a)$ and $\phi_2(\omega_b)$. The R2 PLL results are shown as 2D boxplots for phase lock values of $\phi_2(\omega_a)$ in **Figure 3h** and $\phi_2(\omega_b)$ in **Figure 3i**. The mean values of amplitude, $\overline{A_2}(\omega_a)$ and $\overline{A_2}(\omega_b)$, and frequency, $\overline{\Omega_2}(\omega_a)/2\pi$ and $\overline{\Omega_2}(\omega_b)/2\pi$, are plotted as diamond points in **Figure 3k** (upper), with the color of each point corresponding to the 2D boxplots (**Figure 3h,i**). The corrected phase values, $\phi_2(\omega_a)$ and $\phi_2(\omega_a)$, and mean values of frequency, $\overline{\Omega_2}(\omega_a)/2\pi$ and $\overline{\Omega_2}(\omega_b, t)/2\pi$, are plotted as diamond points in **Figure 3k (lower).** We used these measurements to calculate $\omega_a = \left(\frac{1}{2}\right)(\overline{\Omega_1}(\omega_a) + \overline{\Omega_2}(\omega_a))$ and $\vec{Z}(\omega_a) = \{\overline{A_1}(\omega_a)e^{i\phi_1(\omega_a)}, \overline{A_2}(\omega_a)e^{i\phi_2(\omega_a)}\}$, and similarly to calculate $\omega_b$ and $\vec{Z}(\omega_b)$.

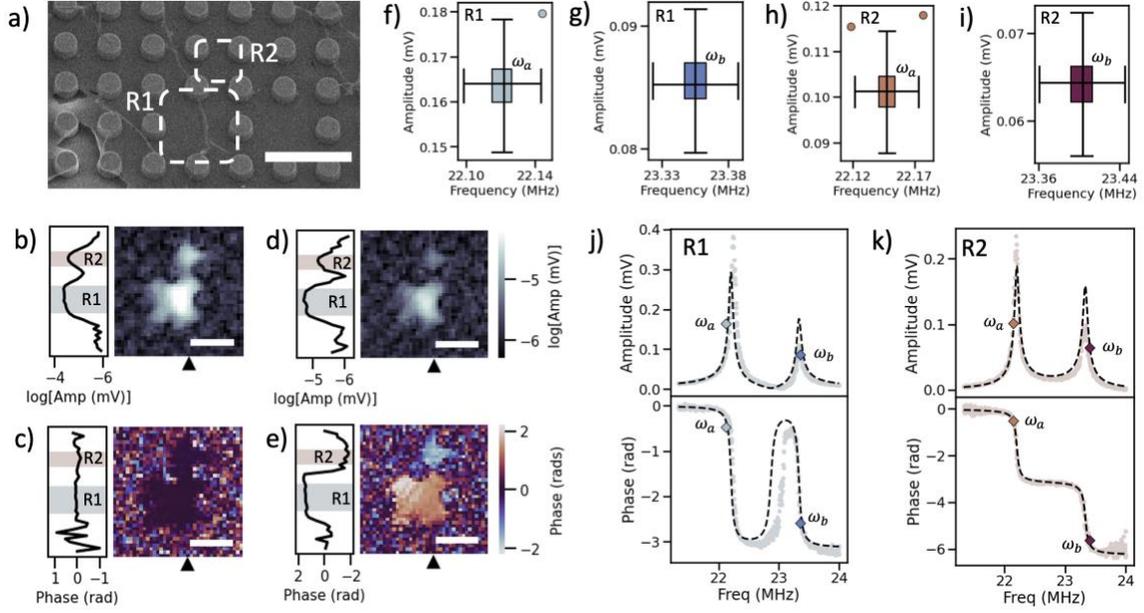

**Figure 3 a)** SEM of driven resonator, R1, and neighboring coupled resonator, R2. Pillar radii are 0.75 μm and pillar pitch is 3 μm, scale bar is 6 μm. **b)** Amplitude and **c)** phase spatial maps taken at a drive frequency of $f_1 =$ 21.51 MHz, scale bars are 5 μm. **d)** Amplitude and **e)** phase spatial maps taken at a drive frequency of $f_2 =$ 23.36 MHz, scale bars are 5 μm. **c)** and **e)** phase spatial maps show uncorrected wrapped phase values. Black triangles on the bottom axes indicate the location of the vertical line scan on the left-side axis. 2D boxplots of PLL measurement distributions of frequency and amplitude for a phase lock of **f)** $\phi_1(\omega_a) = -0.47 \pm 0.06$ rad, **g)** $\phi_1(\omega_b) = -2.59 \pm 0.06$ rad, **h)** $\phi_2(\omega_a) = -0.51 \pm 0.2$ rad, and **i)** $\phi_2(\omega_b) - 5.61 \pm 0.2$. **j)** Amplitude (upper) and corrected phase (lower) of R1. Diamond points in the amplitude plot correspond to PLL measurements of amplitude ($\overline{A_1}(\omega_a) = 0.16368 \pm 0.00005$ mV and $\overline{A_1}(\omega_b) = 0.08560 \pm 0.00003$ mV), and frequency ($\overline{\Omega_1}(\omega_a)/2\pi = 22.1208 \pm 0.00008$ MHz and $\overline{\Omega_1}(\omega_b)/2\pi = 23.3554 \pm 0.0001$ MHz). Diamond points in the phase plot correspond to the phase lock values, $\phi_1(\omega_a)$ and $\phi_1(\omega_b)$, and the mean frequency values, $\overline{\Omega_1}(\omega_a)/2\pi$ and $\overline{\Omega_1}(\omega_b)/2\pi$. The black dotted lines represent $|Z_1(\omega)|$ and $\phi_1(\omega)$ generated from the normalized $\vec{p}$. **k)** Amplitude (upper) and corrected phase (lower) of R2. Diamond points in the amplitude plot (upper) correspond to PLL measurements of amplitude ($\overline{A_2}(\omega_a) = 0.10146 \pm 0.00003$ mV and $\overline{A_2}(\omega_b) = 0.06420 \pm 0.00002$ mV), and frequency ($\overline{\Omega_2}(\omega_a)/2\pi = 22.14721 \pm 0.00007$ MHz and $\overline{\Omega_2}(\omega_b)/2\pi = 23.4029 \pm 0.0001$ MHz). Diamond points in the phase plot (lower) correspond to the phase lock values, $\phi_2(\omega_a)$ and $\phi_2(\omega_b)$, and the mean frequency values, $\overline{\Omega_2}(\omega_a)/2\pi$ and $\overline{\Omega_2}(\omega_b)/2\pi$. The black dotted lines represent $|Z_2(\omega)|$ and $\phi_2(\omega)$ generated from the normalized $\vec{p}$.

To solve for the parameters vector, $\vec{p}$, we populate $\mathcal{Z}$ with coefficients of the experimentally measured values of $\omega_a$, $\omega_b$, $\vec{Z}(\omega_a)$, and $\vec{Z}(\omega_b)$. For a cluster size for $N = 2$, $\vec{p}$ has eight unknown components, $\vec{p} = \{m_1, m_2, b_1, b_2, k_1, k_2, c_1, F\}$ and $\mathcal{Z}$ is an $8 \times 8$ matrix. We applied SVD to solve Eq. (*3*) for $\vec{p}$, with values normalized by $k_1$ listed in **Table 2**. Errors are calculated as described above for the $N = 1$ cluster (see SI). With the $k_1$ normalization, the intrinsic spring $k_2$ is also within the expected range of $1 - 5$ N/m. If we assume R1 and R2 to have equal masses and intrinsic springs, we estimate $c_1 = 0.05$ N/m (see SI), which agrees with the value $c_1$ from **Table 2.** The damping constants $b_1$ and $b_2$ agree within error with the damping predicted above for the $N = 1$ case, which we expect considering that both clusters are on the same sample and tested under the same vacuum[30]. In addition, based on the areas of R1 and R2 and the area density of pristine graphene ($\rho = 0.75$ mg/m$^2$), we predict the mass of each resonator to be $m_1 = 2.7 \times 10^{-17}$ kg and $m_2 = 6.75 \times 10^{-18}$ kg. Accounting for contamination, both $m_1$ and $m_2$ estimates in **Table 2** are plausible. However, we note that NetMAP predicts $m_1 < m_2$, whereas we expect $m_1 > m_2$ because of the relative sizes of R1 and R2. We speculate that differences in local temperatures due to the optical probe position and PLL-related error may account for this mass discrepancy.

| Mechanical Parameter | NetMAP | Unity L-S |
|---|---|---|
| $k_1$ [N/m] | $1 \pm 0.074$ | 1 |
| $k_2$ [N/m] | $1.784 \pm 0.044$ | 1.655 |
| $c_1$ [N/m] | $0.069 \pm 0.002$ | 0.057 |
| $m_1$ [$10^{-17}$ kg] | $5.263 \pm 0.365$ | 5.289 |
| $m_2$ [$10^{-17}$ kg] | $8.975 \pm 0.219$ | 8.165 |
| $b_1$ [$10^{-11}$ kg/s] | $1.424 \pm 1.667$ | 1.530 |
| $b_2$ [$10^{-11}$ kg/s] | $5.405 \pm 4.625$ | 4.846 |
| $F$ [$10^{-6}$ au] | $1.501 \pm 0.651$ | 1.358 |
| R1 Amplitude $R^2$ | 0.73 | 0.89 |
| R1 Phase $R^2$ | 0.83 | 0.99 |
| R2 Amplitude $R^2$ | 0.85 | 0.85 |
| R2 Phase $R^2$ | 0.996 | 0.996 |

**Table 2**: Normalized values of $\vec{p}$ from NetMAP and Unity LS for N=2 cluster

To evaluate the predictive power of $\vec{p}$, we compared the analytical $\vec{Z}_1(\omega)$ and $\vec{Z}_2(\omega)$ generated from $\vec{p}$ to experimental spectra using $R^2$. The analytical responses are shown as black dashed lines overlayed on the spectra in **Figure 3j,k** and the resulting $R^2$ values are listed in **Table 2**. Because the $\vec{p}$-calculated responses in the model account for at least 70% of variability in all the experimental spectra (all $R^2$ values are greater that 0.7), we conclude that the $\vec{p}$-calculated responses can accurately predict the experimental spectral response of the cluster over the tested range.

To further evaluate the utility of our approach, we benchmark the NetMAP predicted $\vec{p}$ against results from Unity LS, see **Table 2**. Although our approach does not require *a priori* information and only uses two vector data points, as opposed to LS which uses thousands of spectral points and requires initial guesses, the two approaches agree but only with precise order-unity *a priori* parameter estimates for LS. Using order-unity guesses informed by the NetMAP $\vec{p}$ (e.g. $m_1 = m_2 = 10^{-17}$ kg) and omitting LS error (see SI), the two approaches agree on the vales $m_1 (p = 0.94)$, $b_1$ ($p = 0.95$), $b_2$ ($p = 0.90$), and $F$ ($p = 0.83$). Although the values disagree for $k_2$ ($p = 0.003$), $c_1$ ($p = 0, t_0 = 7$), and $m_2$ ($p = 0.0002$), each value predicted by LS was still within the expected range discussed above and only differed from the NetMAP $\vec{p}$ by at most $8 - 21\%$. We also found that LS resulted in some higher $R^2$ values, although this is to be expected considering NetMAP does not utilize the full spectra.

The agreement between NetMAP and unity LS is sensitive to the LS input guess solutions. For example, by increasing the input guesses for mass only (e.g. $m_1 = m_2 = 10^{-16}$ kg), LS yields a predicted $k_2$ about $1000 \times$ lower than the expected $1 - 5$ N/m and a predicted $m_2$ about $10 \times$ less than expected for pristine graphene (see SI), as well as poor fits with some $R^2 < 0$. This result highlights the susceptibility of LS to correlation errors because while individually $k_2$ and $m_2$ are far from the expected values, the ratio of $\sqrt{\frac{k_2}{m_2}}$ gives an expected resonance frequency of ~16 MHz, within 7 MHz of the observed amplitude peaks. These correlated values, which are prevalent in our model, result in many possible solutions for least squares fitting, making this method reliant on the input guess solutions to find the residual minimum that is closest to the actual values. Combining NetMAP with LS fitting by inputting $\vec{p}$ as guess solutions may be an effective strategy to fit parameter values as close to the actual values as possible. Given that our approach is not sensitive to input guesses, NetMAP provides a robust means to obtain physically accurate network parameters without any *a priori* knowledge of those parameters.

Thus far, we have reported solutions to $\mathcal{Z}\vec{p} = \vec{0}$ that belong to a one-dimensional (1D) null-space. However, the null-space can be higher dimensional, and considering the full null-space basis may result in a $\vec{p}$ that is more physically accurate. In practice, we characterize the null-space dimension by the number of singular values $\lambda$ such that $\lambda \ll 1$. For the $N = 1$ cluster, (**Figure 2**), we had only one candidate singular value ($\lambda \sim 10^{-9}$), so the null-space was 1D (see SI). However, the $N = 2$ cluster shown in **Figure 3** has three potential singular values ($\lambda_1 \sim 10^{-4}$, $\lambda_2 \sim 10^{-6}$, and $\lambda_3 \sim 10^{-8}$), so the null-space could be up to 3D and the most accurate solution would then be a linear combination of the vectors associated with each singular value, *e.g.* $\vec{p} = \alpha_1 \vec{p}_1 + \alpha_2 \vec{p}_2 + \alpha_3 \vec{p}_3$. In the case of a higher-dimension null-space, we can obtain $\vec{p}$ by specifying additional constraints, such as the estimated masses of each resonator. Alternatively, we could avoid providing *a priori* information about the network parameters by LS fitting to full spectra to solve for the weights of each vector (*i.e.* $\alpha_i$). We applied this LS approach to the $N = 2$ system, and found that both $\vec{p}$ and the corresponding $R^2$ did not change significantly from the **Table 2** values (see SI). However, using the $\alpha_i$ LS fitting for the additional $N = 2$ system (see SI) resulted in the overall best $R^2$ values ($R^2 = 0.94$ for R1 amplitude, $R^2 = 0.98$ for R1 phase, $R^2 = 0.94$ for R2 amplitude, and $R^2 = 0.92$ for R2 phase). Therefore, $\alpha_i$ LS fitting provides a facile means to account for higher-dimensional null-spaces without requiring additional knowledge of network parameters.

The combination of SIM and algebraic solving in the NetMAP approach provides a means to spatially map and quantify the mechanical elements of a resonator network, and as such is a powerful and vital tool for programming resonator networks for future applications.

The spatial information provided by SIM is essential for NetMAP because it provides the size and resonator configuration needed to model the cluster. Since weak coupling is difficult or impossible to detect using the spectra alone[19]—*e.g.* when spectral peaks cannot be resolved because of dissipative broadening—spectroscopy cannot reliably be used to determine model size. Considering this shortfall, SIM is a uniquely useful tool because it detects all resonators with motion in the cluster, regardless of the coupling strength, and therefore quantifies the model size. However, to uniquely define $\mathcal{Z}$ and thereby solve $\vec{p}$, NetMAP requires knowledge of both the cluster size and the exact lateral positions of each resonator in the cluster, including the driven resonator. Spectroscopy alone does not provide spatially resolved information as it cannot be used to distinguish which resonators out of many participate in the hybridized modes. An additional valuable feature of SIM is that it provides the spatial coordinates of all resonators and the force, and thus enables the algebraic characterization method. Moreover, although we have constructed $\mathcal{Z}$ with PLL measurements, it is possible to build $\mathcal{Z}$ directly from SIM amplitude and phase images, where $Z_i(\omega)$ is calculated from spatial averaging over a given resonator. This fully SIM-based route eliminates the need for PLL measurements and isn't susceptible to PLL drift noise but reduces the SNR for the same measurement time.

To program a cluster or entire network of nanomechanical resonators (see **Figure 1a**), it is imperative to know the properties of each resonator node (*i.e.* $k_n, m_n, b_n, c_n$) in its most recent state. As a concrete example, to program a network as a phononic crystal[5,31] the node resonance frequencies ($\omega_n \equiv \sqrt{k_n/m_n}$) and coupling ($c_n$) must be finely tuned. But to adjust $\omega_n$ to a desire value, the current values of $k_n$ and $m_n$ for each resonator must be known before they can be appropriately tuned. This knowledge is especially important when the tuning method is irreversible, such as the additive or subtractive tuning of resonator mass[32]. Although $\omega_n$ can be determined with least squares fitting, correlations in the model make it difficult to decouple the individual values of $k_n$ and $m_n$, and, as discussed above, fitting does not provide spatial information, further obfuscating the node properties. The phononic crystal must also have precisely tuned coupling between resonators. Typically, the coupling strength of simple $N = 2$ systems is inferred from avoided-crossing signatures in gate-tuned spectrographs[14,33]. However, for larger systems (*i.e.* $N \geq 3$) this approach does not directly provide the coupling constants $c_n$ and cannot indicate the corresponding spatial location of the spring analogs. In contrast, NetMAP directly quantifies normalized physical parameters of each node, including $k_n$, $m_n$, and $c_n$, and characterizes the spatial ordering. The normalized values of $\vec{p}$ can be unnormalized by taking a separate measurement of a single parameter, such as measuring $k_n$ with AFM[27]. Therefore, by spatially resolving the network parameters, NetMAP will enable the programming of our network platform for on-demand tailored phononic crystals, engineered defects[34], or control of wave propagation[31].

While we have limited our use of NetMAP to cluster sizes of $N = 1$ and $N = 2$, it can also be used to efficiently map larger clusters and entire networks. For a given cluster of size $N$, the edge coupling spring constants vanish ($c_{n-1} \sim 0$ and $c_{n+N+1} \sim 0$), leaving $4N$ unknown parameters in $\vec{p}$. Regardless of $N$, NetMAP requires a minimum of two response measurements (at $\omega_a$ and $\omega_b$) per resonator to assemble $\mathcal{Z}$, which will be a $4N \times 4N$ matrix. Given the dimensions of $\mathcal{Z}$, the number of SVD-specific operations[35] needed to obtain $\vec{p}$ will scale efficiently with polynomial time, $O(N^3)$. After using NetMAP to characterize each cluster, we can assemble a map of the entire network composed of neighboring clusters separated by coupling springs of $c_n = 0$. In the clustered state, the network will behave like a polycrystalline solid with non-interacting grains. With a full picture of the network, we can modify the edge $c_n$ values to form larger clusters or link clusters into a single, fully interconnected network.

We used NetMAP to characterize one-dimensional suspended graphene resonator networks, but the technique can be applied to higher-dimensional systems and a variety of resonator network analogs. this technique readily scales to networks in two and three dimensions and with more complex coupling beyond nearest-neighbor that we assume in our model. In these cases, depending on the exact geometry of the cluster, it is likely that that more than two complex response vectors must be measured to solve for the unknown parameters. Provided it is possible to obtain spatially resolved amplitude and phase responses for a cluster, as we do here with SIM, our algebraic characterization method can be applied to a range of steady-state linear resonator platforms, such as alternative geometries of MEMS/NEMS arrays[7,36], optical lattices[3], LRC-based electronic circuits, and biological networks. A modified version of NetMAP will also be relevant to quantum analogs, such as chains of trapped ions[1], 2D superconducting qubit arrays[2] and could be useful for programming the initial state for quantum computational tasks.

In conclusion, we use NetMAP to spatially map and quantify all the physical parameters in local suspended graphene resonator clusters of sizes $N = 1$ and $N = 2$. By combining spatially resolved measurements of the resonator building blocks and an algebraic solution of the network equations of motion, NetMAP provides a novel means of characterization that overcomes the weaknesses of current spectroscopic techniques, thereby enabling a programmable resonator network. Programmable NEMS networks will enable applications such as modeling natural systems, realizing mechanical computing schemes[10,20,37,38], integrating into quantum information circuits[39], and exploring new physics such as phononic metamaterials and exotic states[7,9,31].

**Acknowledgements**

The authors thank Rachael Klaiss, Kara Zappitelli, Kurt Langworthy, Rudy Resch, and Jayson Paulose for discussing this work. This work was supported by the Reneé James Seed Grant Initiative, and the National Science Foundation (NSF) under Grant No. CMMI-2128671. The authors acknowledge facilities and staff at the Center for Advanced Materials Characterization in Oregon, and the use of the University of Oregon's Rapid Materials Prototyping facility, funded by the Murdock Charitable Trust.

# Supplementary Information

## Table of Contents

## Discussion 1: Algebraic approach discussion

The linear mass and spring model is described by an infinite set of coupled differential equations.

$$\vdots$$
$$m_{i-1}\ddot{x}_{i-1} + b_{i-1}\dot{x}_{i-1} + k_{i-1}x_{i-1} + c_{i-1}(x_{i-1} - x_i) + c_{i-2}(x_{i-1} - x_{i-2}) = F_{i-1}$$
$$m_i\ddot{x}_i + b_i\dot{x}_i + k_i x_i + c_{i-1}(x_i - x_{i-1}) + c_i(x_i - x_{i+1}) = F_i$$
$$m_{i+1}\ddot{x}_{i+1} + b_{i+1}\dot{x}_{i+1} + k_{i+1}x_{i+1} + c_i(x_{i+1} - x_i) + c_{i+1}(x_{i+1} - x_{i+2}) = F_{i+1}$$
$$\vdots$$

This system of equations can be organized into the matrix form:

$$\mathbf{M}\ddot{\vec{x}} + \mathbf{\Gamma}\dot{\vec{x}} + \mathbf{K}\vec{x} = \vec{F} \tag{5}$$

where the three matrices are

$$M = \begin{pmatrix} \ddots & 0 & 0 & 0 & \iddots \\ 0 & m_{i-1} & 0 & 0 & 0 \\ 0 & 0 & m_i & 0 & 0 \\ 0 & 0 & 0 & m_{i+1} & 0 \\ \iddots & 0 & 0 & 0 & \ddots \end{pmatrix}, \quad \Gamma = \begin{pmatrix} \ddots & 0 & 0 & 0 & \iddots \\ 0 & \gamma_{i-1} & 0 & 0 & 0 \\ 0 & 0 & \gamma_i & 0 & 0 \\ 0 & 0 & 0 & \gamma_{i+1} & 0 \\ \iddots & 0 & 0 & 0 & \ddots \end{pmatrix},$$

$$K = \begin{pmatrix} \ddots & -c_{i-2} & 0 & 0 & \iddots \\ -c_{i-2} & k_{i-1} + c_{i-2} + c_{i-1} & -c_{i-1} & 0 & 0 \\ 0 & -c_{i-1} & k_i + c_{i-1} + c_i & -c_i & 0 \\ 0 & 0 & -c_i & k_{i+1} + c_i + c_{i+1} & -c_{i+1} \\ \iddots & 0 & 0 & -c_{i+1} & \ddots \end{pmatrix}$$

To solve Eq (5), we use a steady-state trial solution of $x_i(t) = Z_i(\omega)e^{i\omega t}$, where $Z_i(\omega) = A_i(\omega)e^{i\phi_i(\omega)}$ is the complex amplitude of the $i^{th}$ resonator. The phase $\phi_i(\omega)$ corresponds to the purely mechanical steady-state phase difference between the driving force and the response $A_i(\omega)$. We then insert this solution into Eq (5) and rewrite the equation as

$$-\omega^2 \mathbf{M}\vec{Z} + i\omega\mathbf{\Gamma}\vec{Z} + \mathbf{K}\vec{Z} = \vec{F} \tag{6}$$

where the two vectors are defined as

$$\vec{Z}(\omega) = \begin{pmatrix} \vdots \\ Z_{i-1}(\omega) \\ Z_i(\omega) \\ Z_{i+1}(\omega) \\ \vdots \end{pmatrix}, \quad \vec{F} = \begin{pmatrix} \vdots \\ F_{i-1} \\ F_i \\ F_{i+1} \\ \vdots \end{pmatrix}$$

More compactly, the equation of motion can be written

$$\boldsymbol{\mathcal{M}}(\omega)\vec{Z}(\omega) = \vec{F} \tag{7}$$

The symmetric, tridiagonal matrix $\boldsymbol{\mathcal{M}}(\omega) = -\omega^2 \mathbf{M} + i\omega\mathbf{\Gamma} + \mathbf{K}$ contains complete information about the inertia, elasticity, and damping of the network. Because the goal of our technique is to calculate the variables in $\boldsymbol{\mathcal{M}}(\omega)$ by measuring $\vec{Z}(\omega)$, we reorganize a finite version of Eq (7) as a homogenous linear equation

$$\boldsymbol{\mathcal{Z}}\vec{p} = \vec{0} \tag{8}$$

In Eq (8), $\vec{p}$ consists of all the unknown mechanical parameters and $\boldsymbol{\mathcal{Z}}$ is a real valued matrix with components calculated from the set drive frequency $\omega$ and the real and imaginary components of $\vec{Z}(\omega)$.

For a cluster of size $N$, the number of unknown parameters in $\vec{p}$ is $4N$; with $N$ masses, $N$ damping constants, $N$ intrinsic springs, $N-1$ coupling springs, and 1 force (assuming one driven resonator). Each measurement of $\vec{Z}(\omega)$ accounts for $2N$ rows in $\boldsymbol{\mathcal{Z}}$, due to the real and imaginary part of each resonator's equation of motion. Therefore, to solve for all the parameters in $\vec{p}$, we will need to take measurements of $\vec{Z}(\omega)$ for a minimum two drive frequencies, such that we have solve $4N$ equations to solve for $4N$ unknowns. We can incorporate measurements taken at additional drive frequencies by using the same SVD package to solve the overdetermined system of equations. In certain cases, the amplitude of some resonators at a given $\omega$ will be "zero", or undetectable by the apparatus. When a resonator's motion is undetectable, the number of linear equations is reduced; if the undetectable resonator is in the interior of the cluster, six equations are lost, while if at the edge of the cluster four equations are lost. Therefore, it is important to ensure that all resonators have a measurable amplitude at the two driving frequencies, or to otherwise take measurements at additional driving frequencies.

## Discussion 2: Fabrication Methods

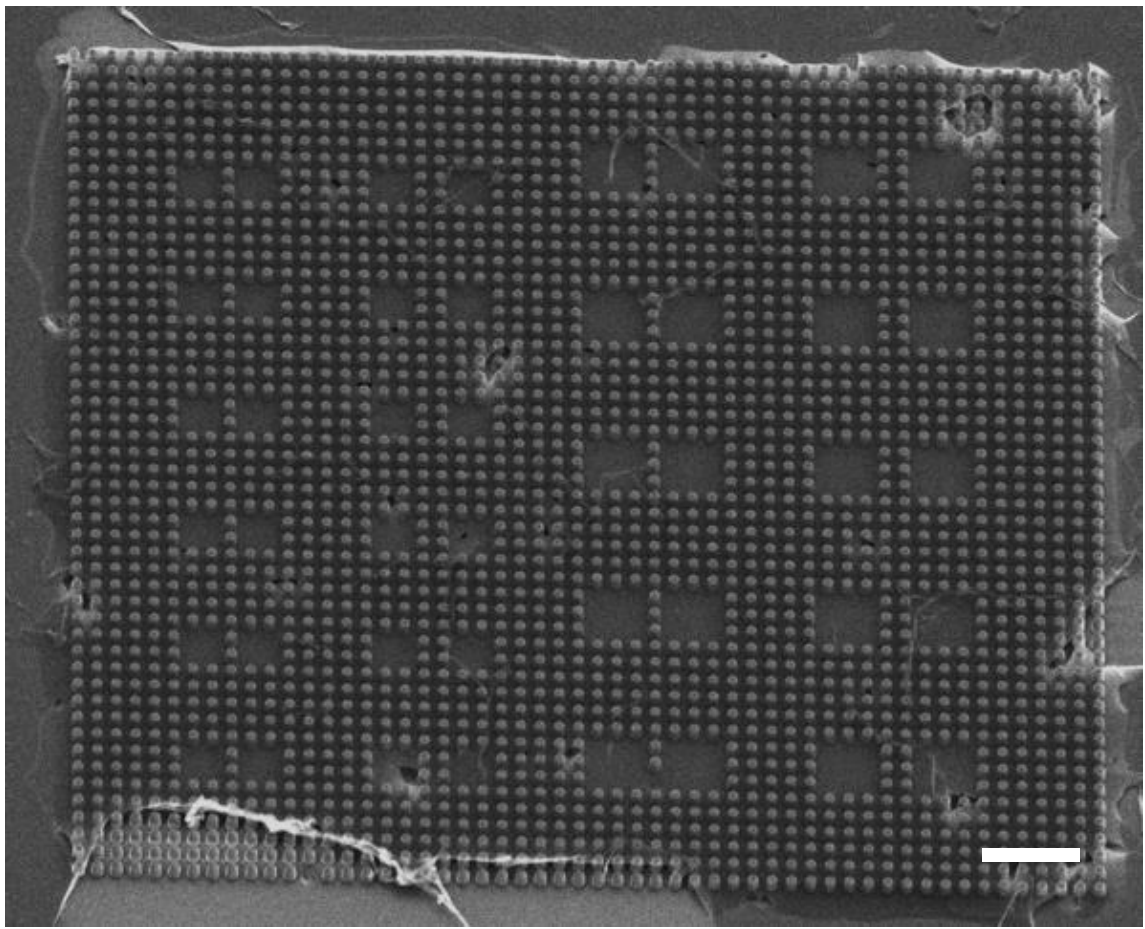

**Figure S1**: Fabrication of suspended graphene resonators. SEM image of full array of suspended graphene over pillars with multiple designated larger resonator pairs. Pillar radius is 0.25 µm and pillar pitch are 1 µm. Scale bar is 5 µm.

    We created the suspended graphene resonator arrays by patterning SiO$_2$/Si substrates using e-beam lithography followed by a dry reactive ion etch (CHF$_4$), resulting in SiO$_2$ pillars $\sim$ 600 nm in height. We began our fabrication process with a doped Si wafer with 1 $\mu$m of commercially grown oxide. To prepare the sample for lithography, we sonicated the diced wafer in acetone, then IPA, followed by a DI rinse and N$_2$ dry. The sample was then dehydrated on a hotplate at 400°C for 30 minutes prior to spinning on PMMA A4 to improve the adhesion of the resist. We used PMMA A4 resist with a thickness of 200nm that was soft baked at 180°C for one minute to evaporate the PMMA solvent. We patterned dot arrays, 0.25 µm - 0.75 µm in radius and 1 µm - 4.5 µm in pitch, into the resist using e-beam lithography with NPGS on a Zeiss scanning electron microscope. In these arrays, we varied the size of the resonators by omitting specific dots and varied the potential coupling strength with 1 or 2 rows of pillars separating the designated resonators. To remove the resist from the exposed dot regions, we developed the sample with MIBK 3:1 for one minute. We then deposited a 20 − 30 nm Cr mask with thermal evaporation in an AMOD followed by a PMMA liftoff in Remover PG on a hotplate at 50°C for one hour. The remaining Cr now

covered only the dot regions that were previously exposed with e-beam lithography. To form the pillars, we used CHF$_4$ in an inductively coupled plasma instrument to etch the exposed regions of SiO$_2$ that were not masked with Cr, resulting in pillars that were $600 \pm 50$ nm. In order to prevent electrical shorting, we left ~400 nm of oxide as an insulating layer, such that any collapsed graphene would not contact the conducting Si directly. Lastly, we etched the remaining Cr, to leave only the SiO$_2$ pillars on silicon. Additionally, we omit pillars throughout the array to create larger-size resonators and resonator pairs (**Figure S1**). The lateral size of the membrane resonators varies from $3 - 6$ µm. The resonators are directly connected by suspended graphene, which provides a mechanism for elastic coupling represented by the coupling spring constants, $c_n$.

We then suspended commercially grown CVD graphene over the pillar arrays using a wet transfer method[1]. To protect the graphene during the transfer, we spun 200 nm of PMMA A4 onto the graphene, resulting in a Cu/Graphene/PMMA sheet. We then used 40 mg/mL ammonium persulfate to etch the copper, leaving only the graphene/PMMA film. We rinsed the graphene/PMMA film in multiple water baths to remove any ammonium persulfate and then used the pillared sample to scoop the film from below, such that the graphene was in direct contact with the sample and the PMMA on top of the graphene. We let the sample dry overnight and then baked the sample at 120°C for 5 minutes to aid the adhesion between the graphene and the substrate. To remove the PMMA, we let the sample soak in acetone for $4 - 6$ hours. To avoid increased tension in the membrane that could cause the membrane to tear, we removed the sample from solution using a critical point dry. The result was a continuous graphene film suspended across pillar arrays to create resonators that could be coupled directly through strain.

## Discussion 3: Experimental Phase lag

Prior to using the experimental phases to populate $\mathcal{Z}$, we first accounted for frequency dependent phase lags caused by time delays of the optical and electronic signal transmission/transduction from the lock-in reference output to the input **(Fig S2a)**. The time-delay phase shift on the pump (i.e the force), $\tau_{force}$, is the sum of delays due to the AOM controller, the AOM, and the free-space optical path and transmission cables up to the sample, shown in **Figure 2Sa**. The total time delay at the lock-in input includes $\tau_{force}$ and $\tau_{response}$, which includes delays due to the photodetector, an "on-chip" thermal lag[2], and the free-space optical path and transmission cables from the device to the lock-in, see **Figure 2Sa**. There is also a constant phase lag due to the phase offset of the lock-in and a $\pi$ phase lag that may arise depending on whether the membrane moves away from or toward the focusing objective[2]. These time delays and offsets will result in a frequency-dependent phase shift according to:

$$\Delta(\omega) = \phi_0 - \omega\tau \qquad (9)$$

Where $\tau$ is the total frequency dependent time delay and $\phi_0$ is the total constant phase offset. The result of this phase lag is evident in the phase data as a linear offset, $\phi_{meas}(\omega)$, shown as light gray or light orange data points in **Figure S2 b-d** upper. We obtain $\tau$ and $\phi_0$ by fitting to a linear section of the raw phase spectra, shown as black solid lines in **Figure S2b,c,d upper.** We then correct for the phase lag by subtracting $\Delta(\omega)$ from the measured phase $\phi_{meas}(\omega)$

$$\phi(\omega) = \phi_{meas}(\omega) - \Delta(\omega) \qquad (10)$$

The corrected phase spectra, $\phi(\omega)$, are shown as light gray or light orange data points in **Figure S2b-d lower.** To account for spatial variations, we perform this calculation for each individual resonator in the cluster. For each measured resonator, we calculated $t_{delay}$ to be ~$700 - 750$ ns, which is consisted with our estimate based on the optical components, lengths of free space optics, and lengths of electrical wiring in our set up. We calculate the propagated error in $\phi(\omega)$ by estimating the uncertainty in the fitted linear section of $\phi_{meas}(\omega)$ using

$$\sigma_\phi = \sqrt{\frac{1}{N-2}\sum_{i=1}^{N}(\phi_i - \phi_0 - \tau\omega_i)^2} \tag{11}$$

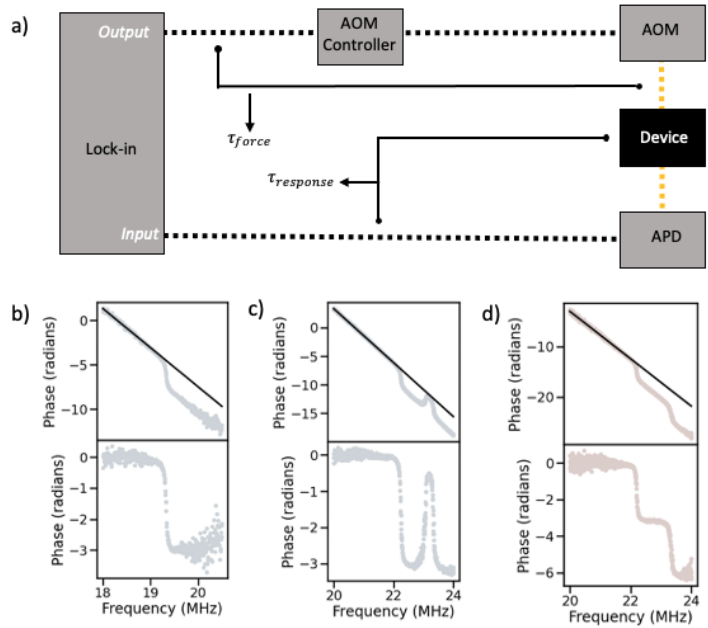

**Figure S2**: Experimental phase lag. **a)** Diagram of the experimental time delay phase shift. Time delay, $\tau_{force}$, includes delays due to transmission cables, shown as black dashed lines, the AOM controller, the AOM, and the free-space optical path, shown as yellow dashed lines. The phase shift on the response, $\tau_{response}$, includes delays due to free-space optical path, the APD, and transmission cables. The total time delay is the sum $\tau = \tau_{force} + \tau_{response}$. **b)** Phase lag correction for $N = 1$ resonator. Upper plot shows uncorrected phase as gray data points and linear fit as black line, with $\tau = 700 \pm 7$ ns and $\phi_0 = 80.5 \pm 0.8$ rad. Lower plot shows corrected phase as gray data points. Phase lag corrections for $N = 2$ for **c)** R1 with upper plot showing uncorrected phase values as gray data points and linear fit as black line, with $\tau = 752 \pm 1$ ns and $\phi_0 = 97.8 \pm 0.1$ rad. Lower plot shows corrected phase values as gray data points and **d)** R2 with upper plot showing uncorrected phase values as light orange data points and linear fit as black line, with $\tau = 747 \pm 4$ ns and $\phi_0 = 90.9 \pm 0.5$ rad. Lower plot shows corrected phase values as light orange data points.

In addition to correcting the phase prior to populating $\mathcal{Z}$, we also considered the experimental lags when choosing target values of $\omega_a$ and $\omega_b$ for PLL. The best choices of $\omega_a$ and $\omega_b$ for high SNR would be at the highest amplitude signals, which occur at peak resonance. However, the peak amplitude at resonance will also correspond to the region of phase with the largest slope. Because we lock into uncorrected phase values, the PLL data will be most affected by the phase lag for values near the steepest slope. We therefore choose target values of $\omega_a$ and $\omega_b$ to be slightly off resonance for the PLL measurements. Future experiments could achieve reduced error with a PLL that accounts for frequency dependent phase lags by locking onto a maximum slope or onto corrected phase values.

## Discussion 4: Errors in parameter vector values

To assess the precision of NetMAP, we solve for $\vec{p}$ with a distribution of $\mathcal{Z}$ matrices and calculate the standard deviation of each output parameter. For the single resonator cluster, we randomly sampled $\omega_i$ and $Z_1(\omega_i)$, for $(i = a, b)$, from normal distributions with a standard deviation equal to the standard error of the PLL time series $\Omega_1(\omega_i, t)$ and $|Z_1(\omega_i, t)|$. We randomly sampled $\phi_1(\omega_i)$ from a normal distribution with the standard deviation equal to the uncertainty calculated in equation Eq (*11*). We then used these randomly sampled values to populate $\mathcal{Z}$ and applied SVD to solve for $\vec{p}$. We repeated this procedure $n = 1000$ times to obtain distributions of each output parameter $(m, b, F)$. For the $N = 1$ cluster, the SVD solver always normalized to the spring constant $k$, so this method did not provide an error for the spring constant. We then removed all solutions that were unphysical or deficient, i.e solutions in which the output parameters did not have a homogenous sign or the $R^2$ values were negative, and were left with distributions of $n = 859$ for each output parameters, with the results shown in **Figure S3a-c**. We quote the standard deviation of each distribution as the error on the parameter value in **Table 1** of the main text. To assess the predictive power of each output parameter set, we used each output $\vec{p}$ to calculate $\vec{Z_n}(\omega)$ and compared the result to the experimental spectra of amplitude and phase to calculate distributions of the $R^2$ values, shown in **Figure S3d**.

We repeated this procedure to calculate the errors for the output parameters of the $N = 2$ cluster. We produced a normal distribution to randomly sample $\omega_i$ with a standard deviation equal to standard errors of the R1 and R2 frequency time series, $\Omega_1(\omega_i, t)$ and $\Omega_2(\omega_i, t)$, added in quadrature. We randomly sampled the amplitudes and phases for each resonator, $Z_1(\omega_i)$ and $Z_2(\omega_i)$, using the same procedure described for the $N = 1$ cluster above. We then use these randomly sampled values to populate $\mathcal{Z}$, used SVD to solve for $\vec{p}$, and repeated the procedure $n = 5000$ times. We then removed all non-physical solutions, resulting in distributions of $n = 253$ shown for each parameter $(m_1, m_2, b_1, b_2, k_1, k_2, k_{12}, F)$ in **Figure S3e-i**. We again assessed the predictive power of each output $\vec{p}$ by comparing calculated $\vec{Z_n}(\omega)$ and to the experimental spectra of the R1 and R2 amplitude and phase to obtain distributions of $R^2$ values, shown in **Figure S3j**.

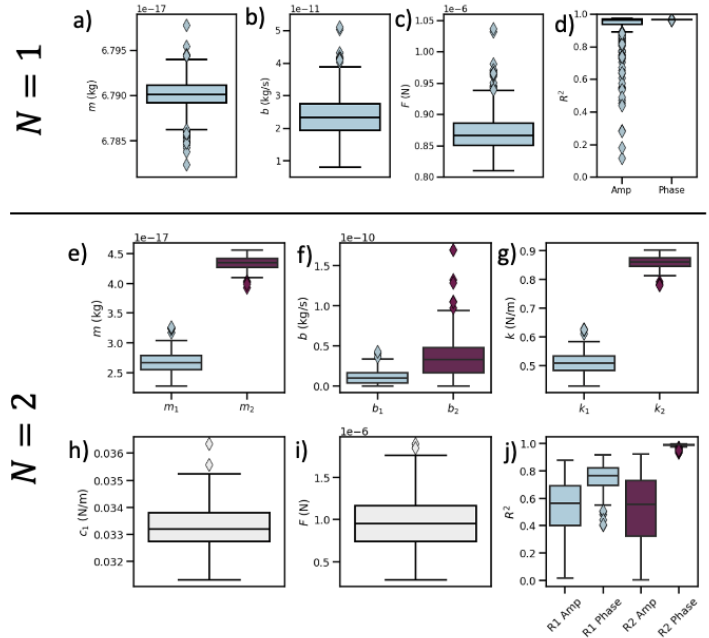

**Figure S3**: Error distributions as box and whisker plots showing median, upper, and lower quartile ranges with whiskers that extend to include 1.5 IQR. Quoted errors are standard deviations of each distribution. Distributions for $N = 1$ cluster parameters shown as light blue boxes with **a)** $m = (6.790 \pm 0.001) \times 10^{-17}$ kg , **b)** $b = (2.4 \pm 0.6) \times 10^{-11}$ kg/s, **c)** $F = (8.7 \pm 0.3) \times 10^{-6}$ N. **d)** $N = 1$ cluster distributions of $R^2$ values for amplitude ($R^2 = 0.94 \pm 0.08$) and phase ($R^2 = 0.968 \pm 0.001$). Distributions for $N = 2$ cluster with R1 distributions shown as light blue boxes and R2 distributions shown as maroon boxes for **e)** $m_1 = (5.2 \pm 0.4) \times 10^{-17}$ kg and $m_2 = (8.5 \pm 0.2) \times 10^{-17}$ kg, **f)** $b_1 = (2.4 \pm 1.9) \times 10^{-11}$ kg/s and $b_2 = (6.7 \pm 4.5) \times 10^{-11}$ kg/s, **g)** $k_1 = 1.00 \pm 0.08$ N/m and $k_2 = 1.69 \pm 0.05$ N/m. Error distributions for $N = 2$ cluster of **h)** $c_1 = 0.065 \pm 0.002$ N/m and **i)** $F = (1.9 \pm 0.7) \times 10^{-6}$ N. **j)** $N = 2$ cluster distributions of $R^2$ values with R1 distributions shown as light blue boxes for amplitude ($R^2 = 0.51 \pm 0.20$) and phase ($R^2 = 0.75 \pm 0.08$). R2 distributions are shown as maroon boxes for amplitude ($R^2 = 0.51 \pm 0.25$) and phase ($R^2 = 0.988 \pm 0.008$).

## Discussion 5: Further details for N=1 cluster in main text

To measure and confirm the size of the the $N = 1$ cluster, we first positioned the pump and probe over the resonator region highlighted in **Figure 2a,** and acquire amplitude and phase spectra. The resulting spectra revealed a single peak in the amplitude, shown as grey data points in **Figure 2f (upper)**, that corresponded to a corrected phase of $\pi/2$ (**Figure 2f (lower)**, grey)**,** consistent with a single, uncoupled resonator. To confirm the size of this cluster was $N = 1$, we took a SIM scan at $18.81 \, MHz$, resulting in the amplitude and phase spatial maps shown in **Figure 2b.** In the amplitude map, we observed an amplitude maximum ($\sim 10^{-4}$ mV) within one localized $\sim 6 \times 6 \, \mu m^2$ region of the suspended graphene, which matched the size and location of the region highlighted in **Figure 2a**. The spatial undulations in the amplitude and phase near the edge of the resonator region are likely due to interactions with the pillars. Outside of the resonator region, the amplitude decreases by more than two orders of magnitude ($\sim 10^{-6}$ mV). Moreover, the resonator region has nearly constant phase ($STD = 0.07$ rad), implying it moves in unison, as expected for the fundamental mode of a single resonator. Away from the resonator, the phase is noisier ($F_0 \sim 74$, $p \sim 10^{-11}$), with an increase in the standard deviation by over an order of magnitude ($STD = 0.6$ rad). Lastly, as seen in the line scans (**Figure 2b,c**), the amplitude has one solitary lobe with constant phase. Altogether, we conclude that this local cluster consists of a single, uncoupled resonator.

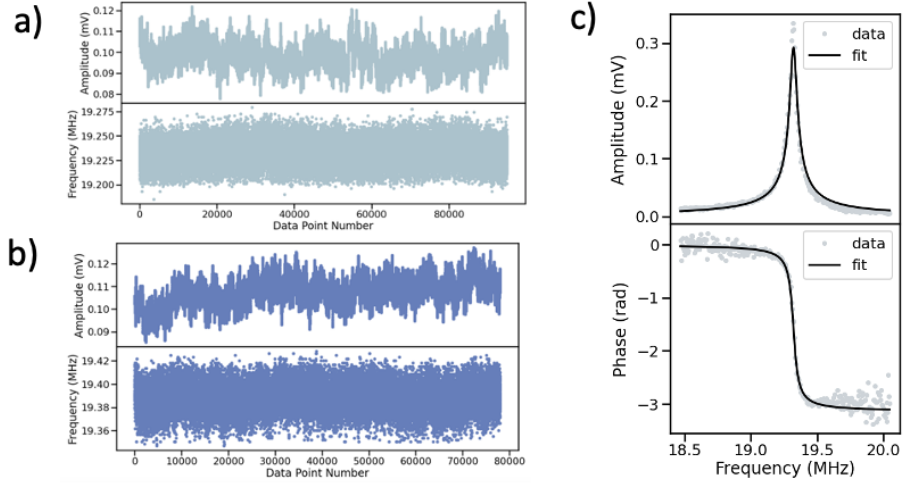

**Figure S4**: Additional figures for $N = 1$ cluster. PLL measurements of amplitude and frequency plotted by data point number for corrected phase lock values of **a)** $\phi_1(\omega_a) = -0.33$ rad shown in light blue and **b)** $\phi_1(\omega_b) = -2.78$ rad shown in dark blue. **c)** least square fit result shown as solid black line over spectra data of amplitude (upper) and phase (lower) shown as gray data points.

To evaluate the stability of the PLL measurements over time, we replot the data that is displayed as 2D boxplots in **Figure 2d,e**, as a function of data point number. The PLL measurements of amplitude (upper) and frequency (lower) are plotted for phase lock values of $\phi_1(\omega_a)$, **Figure S4a**, and $\phi_1(\omega_b)$, **Figure S4b**. We observe that the frequency does not drift significantly in either PLL measurement. We do observe a slight increase in amplitude in the $\phi_1(\omega_b)$ PLL measurement, **Figure S4b** upper, which could be due to a shift up in the resonance frequency due to heating.

In the main text, we evaluate the accuracy of the NETMAP calculated $\vec{p}$ by comparing each calculated value to expected values. To estimate the expected value of the damping, $b$, we fit the amplitude spectra to a built-in Lmfit model, DampedOscillatorModel, which has three fit parameters: $A$, $\omega_0$, and $\sigma$. Because there are low correlations between each parameter, the resulting values have low error. From the model equation and fit parameters $\sigma$ and $\omega_0$, we estimate the damping based on

$$2\sigma\omega_0 = \frac{b}{m}$$

We divide both sides by $\omega_0^2 = \frac{k}{m}$, where $\omega_0$ is the fitted center frequency.

$$b = \frac{2k\sigma}{\omega_0}$$

With fitted values of $\omega_0 = 2\pi \times (19.32 \text{ MHz})$ and $\sigma = 1.16 \times 10^{-3}$, along with the approximated value of $k = 1$ N/m, we estimate the damping to be $b = 1.92 \times 10^{-11}$ kg/s.

To further validate NETMAP, we compare the NETMAP output parameters to those from least-squares fitting with unity order of magnitude guesses, Unity LS. The fit result of Unity LS, listed in **Table 1** in the main text, is shown in **Figure S4c** as a solid black line overlayed on the experimental spectra. To compare each output parameter from NetMAP and LS, we perform a two tailed t-test, using the LS values as reference because the LS errors for each parameter value were large (up to 3000%). We suspect these large errors are due to the high correlations of multiple parameter pairs in our model. We also compared values by using a percent difference, which we calculate by taking the absolute value of the difference between the two values to be compared and dividing by the smaller of the two values for an upper bound estimate.

## Discussion 6: Further details for $N = 2$ cluster in main text

To measure and confirm the size of the $N = 2$ resonator, we positioned the pump and probe over the resonator region highlighted as R1 in **Figure 3a** and acquire amplitude and phase spectra. The resulting amplitude spectrum (**Figure 3f (upper), grey data points**) revealed two closely spaced peaks, each corresponding to a corrected phase of $\sim \pi/2$ (**Figure 3f (lower)**, grey), indicative of the two hybridized modes of a pair of coupled resonators[3]. To test if the spectral features correspond to a resonator pair, we took a SIM scan across the region at a frequency below the first peak ($f_1 = 21.51 \text{ MHz}$). The resulting spatial maps (**Figure 3b,c**) show two distinct high-intensity amplitude regions with nearly constant phase, as highlighted in the line profiles (**Figure 3f**). The first high-amplitude region was centered in a $6 \times 6 \ \mu m^2$ region, which matched the size and location of the driven resonator R1 shown in **Figure 3a**. The second region matched the size and location of a $3 \times 3 \ \mu m^2$ membrane highlighted as R2 in **Figure 3a**. The mean phases of R1 and R2 differed by 0.12 rad, or $\sim 6.9°$, ($p \sim 0.001$), indicating they move in near unison, in accord with expectations for the symmetric mode of a coupled pair[2]. We search for the asymmetric mode with an additional SIM scan at a frequency above the second spectral peak ($f_2 = 22.55 \text{ MHz}$). In the resulting amplitude map and cross-section (**Figure 3d**), we see two regions with high amplitude situated at the same locations as R1 and R2 in **Figure 3a,b**. The phase map and profile (**Figure 3e**) reveal the phase in each region is relatively uniform, but the regions differ from each other by $\sim \pi$ rad, as expected for the asymmetric mode. From the spectra and SIM data, we conclude that the R1 and R2 resonators form a coupled pair.

To further support our conclusion that the spatial maps of amplitude and phase in **Figure 3b-e**, depict a $N = 2$ cluster, we compared the R1 and R2 regions, highlighted in **Figure 3a**, to surrounding areas of low amplitude and noisy phase. In the first SIM scan, **Figure 3b,c**, we observe a decrease in amplitude by more than an order of magnitude outside of the R1 and R2 resonator regions. Moreover, the R1 and R2 regions has nearly constant ($STD = 0.03$ rads, $STD = 0.05$ rads, respectively), whereas the phase outside the resonator regions is noisier ($STD = 0.57$), implying that the R1 and R2 regions each move with constant phase and notable amplitude. For the second SIM scan, **Figure 3d,e**, the amplitude decreased by more than an order of magnitude for R1 and by more than 70% for R2 outside of the two resonator regions. Outside of the resonator regions, the phase was also noisier with about a 70% increase in the standard deviation ($STD = 0.42$).

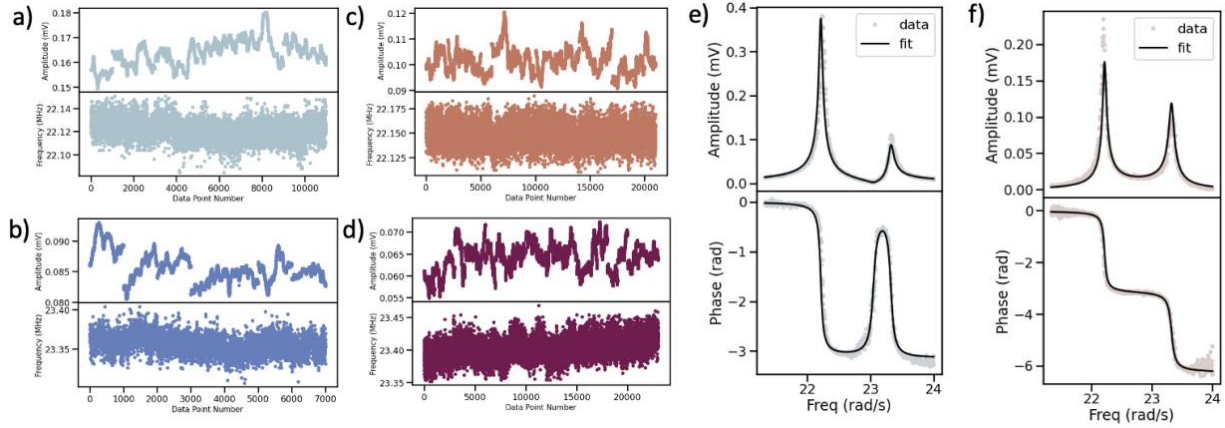

**Figure S5:** Additional figures for $N = 2$ cluster. PLL measurements of amplitude and frequency plotted by data point number for corrected phase lock values of **a)** $\phi_1(\omega_a) = -0.47$ rad shown in light blue, **b)** $\phi_1(\omega_b) = -2.59$ rad shown in dark blue, **c)** $\phi_2(\omega_a) = -0.51$ rad shown in orange, **d)** $\phi_2(\omega_b) = -5.61$ rad shown in maroon. Least squares fit results shown as black solid lines over **e)** R1 amplitude (upper) and phase (lower) shown as gray data points and **f)** R1 amplitude (upper) and phase (lower) shown as light orange data points

To observe the stability of the PLL measurements over time, we replot the data that is displayed as 2D boxplots in **Figure 3f-i**, as a function of data point number. The R1 PLL measurements are shown for phase lock values of $\phi_1(\omega_a)$, **Figure S5a**, and $\phi_1(\omega_b)$, **Figure S5b**. The R2 PLL measurements are shown for phase lock values of $\phi_2(\omega_a)$, **Figure S5c**, and $\phi_2(\omega_b)$, **Figure S5d**. With this representation, we see that the frequency is mostly stable for the $\phi_1(\omega_a)$, $\phi_1(\omega_b)$, and $\phi_2(\omega_a)$ phase locks. However, we do see an increasing trend for the $\phi_2(\omega_b)$ phase lock, which implies that these data are likely affected by the experimental phase lag during the PLL measurement and may account for error in the calculation of $\vec{p}$.

In the main text, we evaluate the accuracy of the NETMAP calculated $\vec{p}$ by comparing each calculated value to expected values. To estimate the expected value of the coupling strength, we start with the equation[3]

$$\Delta\omega = \sqrt{\frac{(c/m_1)(c/m_2)}{\omega_1 \omega_2}}$$

In this equation, $\omega_1 = \sqrt{\frac{k_1+c}{m_1}}$, $\omega_2 = \sqrt{\frac{k_2+c}{m_2}}$, and $\Delta\omega$ is approximated to be the difference between the antisymmetric mode frequency and the symmetric mode frequency, $\Delta\omega = \omega_A - \omega_S$. We simplify this equation by assuming the two resonators have equal mass, $m$, and equal intrinsic spring constants, $k$. We then normalize by the eigenfrequency of the uncoupled resonators, $\omega_0 = \sqrt{\frac{k}{m}}$.

$$\frac{\Delta\omega}{\omega_0} = c\sqrt{\frac{1}{k(k+c)}}$$

We can then solve for $c$, in which we define $\Omega = \frac{\Delta\omega}{\omega_0}$,

$$c = \frac{1}{2}\left(k\Omega^2 \pm \sqrt{(k\Omega^2)^2 + (2k\Omega)^2}\right)$$

We used the SciPy find_peaks function in Python to estimate $\Delta\omega = 1.1$ MHz and $\omega_0 = \omega_S + \frac{\Delta\omega}{2} = 22.8$ MHz. Using these values and approximating $k = 1$ N/m, we obtain an estimate of $c = 0.05$ N/m.

To further validate NETMAP, we compare the NETMAP output parameters to those from Unity LS. The fit result of Unity LS, used for comparison in the main text, is shown as a solid black line overlayed on the experimental spectra, gray or light orange data points, in **Figure S5e,f**.

To highlight the sensitivity of LS to input guesses, we also performed LS with increasing our input guess of the two masses from $m_1 = m_2 = 10^{-17}$ kg to $10^{-16}$ kg. For this fit, we saw order of magnitude or more differences in the output values of $k_2$, $c_1$, $m_2$, $b_2$, and $F$ that do not fall within the expected ranges discussed in the main text. While individually $k_2$ and $m_2$ are far from the expected values, the ratio of $\sqrt{\frac{k_2}{m_2}}$ gives an expected resonance frequency of ~16 MHz, within 7 MHz of the observed amplitude peaks. These correlated quantities, which are prevalent in our model, result in many possible solutions for least squares fitting, making this method reliant on the input guess solutions to find the residual minimum that is closest to the actual values. Combining NetMAP with LS fitting by inputting $\vec{p}$ as guess solutions (see SI for resulting $\vec{p}$ and $R^2$ values) may be an effective strategy to fit parameter values as close to the actual values as possible.

As a final comparison, we used LS to fit the vector weights of the two smallest singular values. In practice, we characterize the null-space dimension by the number of singular values $\lambda$ such that $\lambda \ll 1$. For the $N = 1$ cluster, (**Figure 2**), we had only one singular value ($\lambda \sim 10^{-9}$), so the null-space was 1D. However, the $N = 2$ cluster (**Figure 3**) had three potential singular values ($\lambda_1 \sim 10^{-4}$, $\lambda_2 \sim 10^{-6}$, and $\lambda_3 \sim 10^{-8}$). We applied the $\alpha_i$ LS approach to the two smallest singular values of the $N = 2$ system and found that all output values still fall within the expected ranges and the corresponding $R^2$ did not change significantly.

| Mechanical Parameter | LS with guess of $m_1 = m_2 = 10^{-16}$ kg | $\alpha_i$ LS |
|---|---|---|
| $k_1$ [N/m] | 1 | 1 |
| $k_2$ [N/m] | 0.004 | 1.686 |
| $c_1$ [N/m] | 0.004 | 0.065 |
| $m_1 \times 10^{-17}$ [kg] | 5.022 | 5.247 |
| $m_2 \times 10^{-17}$ [kg] | 0.038 | 8.481 |
| $b_1 \times 10^{-11}$ [kg/s] | 1.452 | 2.250 |
| $b_2 \times 10^{-11}$ [kg/s] | 0.022 | 5.149 |
| $F \times 10^{-6}$ [N] | 0.112 | 1.790 |
| R1 Amplitude $R^2$ | −0.27 | 0.79 |
| R1 Phase $R^2$ | 0.99 | 0.84 |
| R2 Amplitude $R^2$ | 0.78 | 0.67 |
| R2 Phase $R^2$ | 0.996 | 0.995 |

**Table S3**: LS with alternate guess solutions, LS to solve for vector weights of $N = 2$ cluster.

# Discussion 7: Additional Cluster Analysis ($N = 1$ and $N = 2$)

To further demonstrate NetMAP, we characterized additional clusters of size $N = 1$ and $N = 2$. We first characterized the $N = 1$ cluster by aligning the pump and probe over the resonator highlighted in **Figure S6a** and acquired amplitude and phase spectra. The resulting spectra had a single peak in the amplitude, gray data points in **Figure S6f upper**, and a corresponding phase that crossed through $\pi/2$ at resonance, gray data points in **Figure S6f lower**, consistent with a single, uncoupled resonator. To confirm the cluster size of $N = 1$, we took a SIM scan at $f = 17.64$ MHz, with the resulting amplitude spatial map shown in **Figure S6b** and phase in **Figure S6c**. In this scan we observed a single region of peak amplitude that had a constant phase, indicating that the cluster consisted of a single, uncoupled resonator. This region also corresponded in size and location to the driven resonator highlighted in **Figure S6a**.

We next measured $\omega_a, \omega_b, \vec{Z}(\omega_a)$ and $\vec{Z}(\omega_b)$, which are needed to compute $\mathcal{Z}$. We chose target values of $\omega_a$ and $\omega_b$ to be on either side of the resonance peak, **Figure S6f (upper)**. We used the corrected phase spectra, **Figure S6f (lower)**, to map the chosen values of $\omega_a$ and $\omega_b$ to phase values for the PLL. The measured PLL time series of amplitude, $A(\omega_{a,b})$, and frequency, $\Omega(\omega_{a,b})/2\pi$, are shown as 2D boxplots in **Figure S6d,e**. Using the mean values of amplitude, $\overline{A}(\omega_a)$ and $\overline{A}(\omega_b)$, and frequency, $\overline{\Omega}(\omega_a)$ and $\overline{\Omega}(\omega_b)$, we obtain $\omega_a = \overline{\Omega}(\omega_a)$, $\omega_b = \overline{\Omega}(\omega_b)$, $\vec{Z}(\omega_a) = \{\overline{A}(\omega_a)e^{i\phi(\omega_a)}\}$, and $\vec{Z}(\omega_b) = \{\overline{A}(\omega_b)e^{i\phi(\omega_b)}\}$.

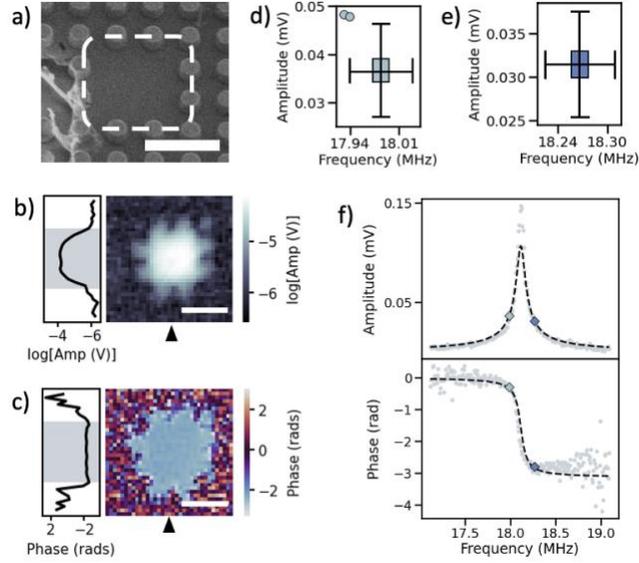

**Figure S6**: Additional $N = 1$ cluster **a)** SEM of driven uncoupled resonator, with pillar radii of 0.5 µm and pillar pitch 2 µm. Scale bar is 4 µm. **b)** Amplitude and **c)** phase spatial maps at a drive frequency of 17.64 MHz, scale bars are 5 µm. Phase spatial map shows uncorrected wrapped phase values. Black triangles on the bottom axis indicate the location of the vertical line scan on the left-side axis. 2D boxplot of PLL measurement distribution of frequency and amplitude for corrected phase lock values of **d)** $\phi_1(\omega_a) = -0.30$ rad and **e)** $\phi_2(\omega_b) = -2.79$ rad. 2D boxplots shows median, upper, and lower quartile ranges with whiskers that extend to include 1.5 IQR. Plotted circles represent datapoints that were outliers in both frequency and amplitude. **f)** Amplitude and corrected phase spectra of driven resonator. The diamond points in the amplitude spectrum (upper) correspond to the mean PLL measurements of amplitude ($\overline{A_1}(\omega_a) = 0.03684 \pm 0.00001$ mV and $\overline{A_1}(\omega_b) = 0.03146 \pm 0.00001$ mV) and frequency ($\overline{\Omega_1}(\omega_a)/2\pi = 17.98422 \pm 0.00005$ MHz and $\overline{\Omega_1}(\omega_b)/2\pi = 18.26637 \pm 0.00007$ MHz). The diamond points in the phase spectrum (lower) correspond to the locked phase values, $\phi_1(\omega_a)$ and $\phi_1(\omega_b)$, and the mean frequency values, $\overline{\Omega_1}(\omega_a)/2\pi$ and $\overline{\Omega_1}(\omega_b)/2\pi$. The black dotted line represents $|Z_1(\omega)|$ and $\phi_1(\omega)$ generated from the normalized $\vec{p}$.

We then populated the matrix $\mathcal{Z}$ with coefficients of the experimentally measured $\omega_a, \omega_b, \vec{Z}(\omega_a)$ and $\vec{Z}(\omega_b)$ and solved for the parameters vector $\vec{p}$. The resulting values are normalized to $k$ and listed in **Table S4**. We found that each predicted parameter in $\vec{p}$ was consistent with the expected ranges discussed in the main paper. Moreover, by comparing the analytical $\vec{Z}_1(\omega)$ and $\vec{Z}_2(\omega)$ generated from $\vec{p}$ to experimental spectra, we found that the model can account for 95% of the variation in the data, with $R^2$ values listed in **Table S4**. We also compared the results from NetMAP to those from Unity LS and found that $m$ ($p = 0.35$), $b$ ($p = 0.91$), and $F$ ($p = 0.41$) all agree. We therefore conclude that NetMAP is proficient in characterizing the local cluster.

| Mechanical Parameter | NetMAP | Unity LS |
|---|---|---|
| $k$ [N/m] | 1 | 1 |
| $m \times 10^{-17}$ [kg] | $7.718 \pm 0.010$ | 7.709 |
| $b \times 10^{-11}$ [kg/s] | $4.595 \pm 2.814$ | 4.293 |
| $F \times 10^{-7}$ [N] | $5.581 \pm 0.527$ | 6.017 |
| Amplitude $R^2$ | 0.94 | 0.95 |
| Phase $R^2$ | 0.97 | 0.97 |

**Table S4**: Results for additional $N = 1$ cluster from NetMAP and Unity LS.

We characterized the additional $N = 2$ cluster by first aligning the pump and probe over the region highlighted as R1 in **Figure S7a** and taking amplitude and phase spectra. The amplitude spectrum, gray data points in **Figure S7j**, revealed two peaks that corresponded to phase changes that pass through ~$\pi/2$ on resonance. To test whether these peaks correspond to the hybridized modes of a coupled pair, we took an SIM scan at $f = 15.45$ MHz, **Figure S7b,c**, and another at $f = 16.21$ MHz, **Figure S7d,e**. In the first scan, we saw two distinct high intensity area regions with nearly constant phase across each individual resonator. The first $6 \times 6$ µm region matched the size and location of the driven resonator, R1 highlighted in **Figure S7a**. The neighboring $6 \times 6$ µm region matched the size and location of the resonator to the right of R1, labelled as R2 in **Figure S7a**. Although we noticed smaller neighboring regions of amplitude maxima and constant phase, because we only observe two peaks in the amplitude spectra, we approximate other membrane motion to be due to weak coupling. We therefore approximate this system to be an $N = 2$ cluster.

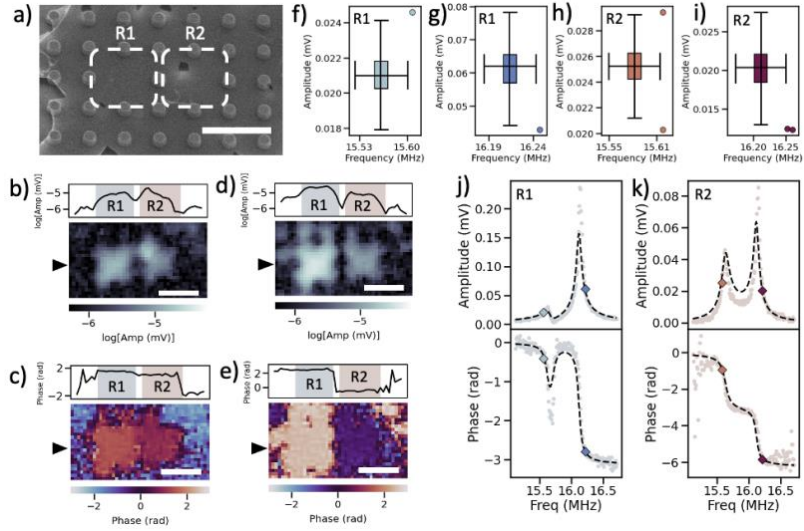

**Figure S7**: Additional $N = 2$ cluster. **a)** SEM of driven resonator, R1, and neighboring coupled resonator, R2. Pillar radii are 0.5 μm and pillar pitch is 3 μm, scale bar is 6 μm. **b)** Amplitude and **c)** phase spatial maps taken at a drive frequency of $f_1 = 15.45$ MHz, scale bars are 5 μm. **d)** Amplitude and **e)** phase spatial maps taken at a drive frequency of $f_2 = 16.21$ MHz, scale bars are 5 μm. **c)** and **e)** phase spatial maps show uncorrected wrapped phase values. Black triangles on the left-side axes indicate the location of the vertical line scan on the left-side axis. 2D boxplots of PLL measurement distributions of frequency and amplitude for a phase lock of **f)** $\phi_1(\omega_a) = -0.41 \pm 0.15$ rad, g) $\phi_1(\omega_b) = -2.79 \pm 0.15$ rad, h) $\phi_2(\omega_a) = -0.93 \pm 0.81$ rad, and **i)** $\phi_2(\omega_b) = -5.85 \pm 0.81$. j) Amplitude (upper) and corrected phase (lower) of R1. Diamond points in the amplitude plot correspond to PLL measurements of amplitude ($\overline{A_1}(\omega_a) = 0.021010 \pm 0.000007$ mV and $\overline{A_1}(\omega_b) = 0.060903 \pm 0.000020$ mV), and frequency ($\overline{\Omega_1}(\omega_a) = 15.56192 \pm 0.00009$ MHz and $\overline{\Omega_1}(\omega_b) = 16.21389 \pm 0.0004$ MHz). Diamond points in the phase plot correspond to the phase lock values, $\phi_1(\omega_a)$ and $\phi_1(\omega_b)$, and the mean frequency values, $\overline{\Omega_1}(\omega_a)$ and $\overline{\Omega_1}(\omega_b)$. The black dotted lines represent $|Z_1(\omega)|$ and $\phi_1(\omega)$ generated from the normalized $\vec{p}$. **k)** Amplitude (upper) and corrected phase (lower) of R2. Diamond points in the amplitude plot (upper) correspond to PLL measurements of amplitude ($\overline{A_2}(\omega_a) = 0.025206 \pm 0.000008$ mV and $\overline{A_2}(\omega_b) = 0.020166 \pm 0.000007$ mV), and frequency ($\overline{\Omega_2}(\omega_a) = 15.58158 \pm 0.00007$ MHz and $\overline{\Omega_2}(\omega_b) = 16.21161 \pm 0.00004$ MHz). Diamond points in the phase plot (lower) correspond to the phase lock values, $\phi_2(\omega_a)$ and $\phi_2(\omega_b)$, and the mean frequency values, $\overline{\Omega_2}(\omega_a)$ and $\overline{\Omega_2}(\omega_b)$. The black dotted lines represent $|Z_2(\omega)|$ and $\phi_2(\omega)$ generated from the normalized $\vec{p}$.

To measure $\omega_a, \omega_b, \vec{Z}(\omega_a)$ and $\vec{Z}(\omega_b)$, we chose a target value of $\omega_a$ below the symmetric mode peak and $\omega_b$ above the antisymmetric mode peak, shown for R1 in **Figure S7j (upper)** and for R2 in **Figure S7k (upper)**. We used the corrected R1 phase spectra, **Figure S7j (lower)**, to map the chosen $\omega_a$ and $\omega_b$ to phase values of $\phi_1(\omega_a)$ and $\phi_1(\omega_b)$. Because the R2 phase spectrum off resonance was too noisy to accurately fit to a linear model, we corrected the R2 phase values with the R1 fit values of $\phi_0$ and $\tau$. With the probe positioned over R1, we acquired PLL time-series measurements for the two phase values, shown as 2D boxplots in **Figure S7f,g**. We repeated these PLL measurements for the second resonator by using the corrected R2 phase spectra, **Figure S7k (lower)**, to map $\omega_a$ and $\omega_b$ to phase values of $\phi_2(\omega_a)$ and $\phi_2(\omega_b)$. We positioned the probe over R2, fixed the pump over R1, and acquired PLL time-series measurements for each phase, shown as 2D boxplots **Figure S7h,i**. We then used the PLL measurements to calculate $\omega_a = \left(\frac{1}{2}\right)(\overline{\Omega_1}(\omega_a) + \overline{\Omega_2}(\omega_a))$, $\vec{Z}(\omega_a) = \{\overline{A_1}(\omega_a)e^{i\phi_1(\omega_a)}, \overline{A_2}(\omega_a)e^{i\phi_2(\omega_a)}\}$, $\omega_b = \left(\frac{1}{2}\right)(\overline{\Omega_1}(\omega_b) + \overline{\Omega_2}(\omega_b))$, and $\vec{Z}(\omega_b) = \{\overline{A_1}(\omega_b)e^{i\phi_1(\omega_b)}, \overline{A_2}(\omega_b)e^{i\phi_2(\omega_b)}\}$.

Using the calculated values of $\omega_a$, $\omega_b$, $\vec{Z}(\omega_a)$, and $\vec{Z}(\omega_b)$, we populated the matrix $\mathcal{Z}$ and solved for the parameters vector $\vec{p}$. We applied SVD to solve $\mathcal{Z}\vec{p} = \vec{0}$ for the eight unknown components of $\vec{p}$, which are normalized by $k_1$ and listed in **Table S5**. We find that the values of $k_2, c_1, m_1, m_2, b_1$, and $b_2$ are all within the expected ranges discussed in the main text. Moreover, we observe that this set of parameters has predictive power over the experimentally measured spectral range, with R1 and R2 amplitude $R^2$ values $\geq 0.77$ and the phase $R^2$ values $\geq 0.92$ (See **Table S5**).

| Mechanical Parameter | NetMAP | Unity LS | $\alpha_i$ LS |
|---|---|---|---|
| $k_1$ [N/m] | $1 \pm 0.163$ | 1 | 1 |
| $k_2$ [N/m] | $0.923 \pm 0.299$ | 0.664 | 1.301 |
| $c_1$ [N/m] | $0.021 \pm 0.006$ | 0.021 | 0.029 |
| $m_1 \times 10^{-17}$ [kg] | $10.035 \pm 1.544$ | 10.039 | 10.113 |
| $m_2 \times 10^{-17}$ [kg] | $9.695 \pm 3.134$ | 7.032 | 13.661 |
| $b_1 \times 10^{-11}$ [kg/s] | $4.336 \pm 4.00$ | 4.476 | 2.650 |
| $b_2 \times 10^{-11}$ [kg/s] | $7.396 \pm 5.546$ | 4.311 | 10.204 |
| $F \times 10^{-6}$ [N] | $0.883 \pm 0.355$ | 0.910 | 0.736 |
| R1 Amplitude $R^2$ | 0.88 | 0.90 | 0.94 |
| R1 Phase $R^2$ | 0.97 | 0.98 | 0.98 |
| R2 Amplitude $R^2$ | 0.77 | 0.35 | 0.94 |
| R2 Phase $R^2$ | 0.92 | 0.93 | 0.92 |

**Table S5**: Results for additional $N = 2$ cluster

We also benchmarked NetMAP results against those form least squares fitting. In comparing the values from NetMAP to those of Unity LS, we find that all values agree within error for $k_2$ ($p = 0.38$), $c_1$ ($p = 0.99$), $m_1$ ($p = 0.998$), $m_2$ ($p = 0.40$), $b_1$ ($p = 0.97$), $b_2$ ($p = 0.58$), and $F$ ($p = 0.94$). However, the predicted power of the Unity LS is poor for the R2 amplitude, as the model only accounts for 35% of variation in the experimental data (See **Table S5**). However, the method that had the best predictive power over the tested experimental range, was when we used LS to fit the vector weights corresponding to the two smallest singular values, with R1 and R2 amplitude and phase $R^2$ values all $\geq 0.92$.

## Supplementary Information References